\documentclass[iop]{emulateapj}
\usepackage{hyperref}
\usepackage{amsmath}

\def\ne{N(C$^{18}$O)\ }
\def\nthir{N($^{13}$CO)\ }
\def\nthirns{N($^{13}$CO)}
\def\ceio{C$^{18}$O\ }
\def\ceions{C$^{18}$O}
\def\co{$^{12}$CO\ }
\def\cons{$^{12}$CO}
\def\thirco{$^{13}$CO\ }
\def\thircons{$^{13}$CO}
\def\av{A$_{\rm V}$\ }
\def\avns{A$_{\rm V}$}
\def\tex{T$_{\rm ex}$\ }
\def\texns{T$_{\rm ex}$}
\def\tk{T$_{\rm k}$\ }

\def\tbg{T$_{\rm bg}$\ }
\def\tmb{T$_{\rm mb}$\ }
\def\tmbns{T$_{\rm mb}$}
\def\td{T$_{\rm d}$\ }
\def\xco{X$_{\rm CO}$\ }
\def\xcons{X$_{\rm CO}$}

\shorttitle{Dust and Gas CO in CMC}
\shortauthors{Kong et al.}

\begin{document}

\title{The Relationship Between the Dust and Gas-Phase CO Across the California Molecular Cloud}

\author{S. Kong\altaffilmark{1}}
\affil{Astronomy Department, University of Florida, FL 32611}
\email{skong@astro.ufl.edu}

\author{C. J. Lada\altaffilmark{2}}
\affil{Harvard-Smithsonian Center for Astrophysics, 60 Garden Street, Cambridge, MA 02138}

\author{E. A. Lada\altaffilmark{1}}
\affil{Astronomy Department, University of Florida}

\author{C. Rom\'an-Z\'u\~niga\altaffilmark{3}}
\affil{Instituto de Astronom\'ia, Universidad Nacional Aut\'onoma de M\'exico, Unidad Acad\'emica en Ensenada, Km 103 Carr. Tijuana-Ensenada, 22860 Ensenada BC, Mexico}

\author{J. H. Bieging\altaffilmark{4}}
\affil{Steward Observatory, The University of Arizona, Tucson, AZ 85719}

\author{M. Lombardi\altaffilmark{5}}
\affil{Department of Physics, University of Milan, Italy}

\author{J. Forbrich\altaffilmark{6} \& J. F. Alves\altaffilmark{6}}
\affil{University of Vienna, Vienna, Austria}

\begin{abstract}
We present results of an extinction--CO line survey 
of the southeastern part of the California Molecular Cloud (CMC). 
Deep, wide-field, near-infrared images were 
used to construct a sensitive, 
relatively high resolution ( $\sim$ 0.5 arc  min)  (NICEST) extinction 
map of the region. The same region was also surveyed in 
the \cons(2-1), \thircons(2-1), \ceions(2-1) emission 
lines at the same angular resolution.
These data were used to investigate the relation between 
the molecular gas, traced by CO emission lines, and the dust 
column density, traced by extinction, on spatial scales of 0.04 pc 
across the cloud. We found strong spatial variations in the abundances 
of \thirco and \ceio that were correlated with variations in 
gas temperature, consistent with temperature dependent CO depletion/desorption
on dust grains. The \thirco to \ceio abundance ratio was found to increase
with decreasing extinction, suggesting selective photodissociation of \ceio  by the 
ambient UV radiation field. The effect is particularly pronounced 
in the vicinity of an embedded cluster where the UV radiation appears to have penetrated deeply
(i.e., \av $\la$ 15 mag) into the cloud. 
We derived the cloud averaged X-factor to be $<$\xcons$>$ $=$ 
2.53 $\times$ 10$^{20}$~${\rm cm}^{-2}~({\rm K~km~s}^{-1})^{-1}$, 
a value somewhat higher than the Milky Way average.
On sub-parsec scales we find there is no single empirical value of the \co
X-factor that can characterize the molecular gas in cold (\tk $\la$ 15 K) cloud regions, with 
\xco $\propto$ A$_{\rm V}^{0.74}$ for \av $\ga$ 3 magnitudes. 
However in regions containing relatively hot (\tex $\ga$ 25 K) 
molecular gas we find a clear correlation between W(\cons) 
and \av  over a large (3 $\la$ \av $\la$ 25 mag) range of extinction. 
This results in a constant \xco $=$ 1.5 $\times$ 10$^{20}$
${\rm cm}^{-2}~({\rm K~km~s}^{-1})^{-1}$ for the hot gas,
a  lower value than either the average for the CMC or the Milky Way.
Overall we find an (inverse) correlation between \xco and \tex in the 
cloud with \xco $\propto$ \texns$^{-0.7}$.  This correlation suggests that 
the global X-factor of a Giant Molecular Cloud (GMC) may depend on 
the relative amounts of hot gas contained within the cloud. 
\end{abstract}


\section{Introduction}

The relationship between the dust and gas in a molecular cloud
is crucial in understanding the cloud properties and the star
formation activities therein. 
Dust provides the most reliable and useful tracer of the total 
gas column density \citep[e.g.,][]{bohlin78,rachford2009,2009ApJ...692...91G}, 
while molecules like \thirco and \ceio
are more useful as tracers of the local physical and chemical conditions (e.g., gas temperatures and velocities, CO column densities, etc.)
A comparison between them allows us to probe the variation of
molecular gas abundances, i.e.,  [\thircons] and [\ceions]
([molecule] $\equiv$ N(molecule)/N(H) where N is column density),
caused by such factors as photodissociation and chemical 
depletion/desorption in cold/warm environments.
This helps to reveal the physical and chemical properties
of the molecular cloud, which enables us to better understand the initial conditions
of star formation and its potential dependence on such factors as cloud chemistry and structure.
Such variations have been noted in previous observations of a variety of GMCs, including 
Perseus \citep[]{pineda08}, Taurus \citep[]{pineda10}, and  Orion \citep{ripple13}, 
made at relatively high spatial resolutions (0.2 - 0.4 pc).
Here we report the results of a combined CO-dust extinction 
study of the California molecular cloud \citep[CMC, ][]{lada09} 
at even higher spatial resolution ( $\sim$ 0.04 pc) in order to 
investigate in more detail the relation between CO gas and 
dust in this interesting nearby GMC.

The CMC is a useful laboratory for such studies. In particular, 
it contains a massive filamentary structure in its southeast region that consists of both 
cold molecular gas with little star formation activity and 
hot molecular gas associated with an active star forming region 
containing the massive star LkH$\alpha$ 101 and its accompanying embedded cluster. 
This provides a physical environment with large temperature 
and density variations that are interesting to explore. 
Indeed, as reported in  \citet{lada09}, the CMC overall shows 
an order of magnitude lower star formation rate (SFR) than the 
Orion molecular cloud, even though these equally distant GMCs 
have similar mass, size, and morphology. Recently, \citet{2014A&A...567A..10L} 
also found the \thirco clumps in the CMC to be characterized by 
similar kinematic states as those in Orion. \citet{lada09} proposed 
that the difference in SFRs between the two clouds was a result of a 
difference in cloud structure, specifically the amount of dense gas in each cloud. Moreover,
this close connection between dense gas mass and the SFR appears to 
be a general physical property of galactic GMCs \citep{2010ApJ...724..687L}. 
It would be of interest to compare other properties of the CMC with similar 
ones in the Orion cloud to look for additional factors, such as cloud 
chemistry for example, that could contribute to the difference in the SFR between the two clouds. 

A combined  CO-extinction study can also be used to investigate the CO X-factor
(\xco $\equiv$ N(H$_2$)/W$_{\rm CO(1-0)}$, where W is integrated intensity) 
which is widely used to derive molecular cloud masses from the CO(1-0) line flux,
especially in external galaxies \citep[see e.g.][]{2013ApJ...777....5S,bolatto2013}.
Despite being very optically thick, this CO rotational transition is still thought
to be a good tracer for the total molecular gas mass, 
due to the effect of photon transport in gas with large velocity gradients. 
However, variations of physical conditions can have an effect on the X-factor and, 
in many instances, render its application rather uncertain. 
Apparently, there is no universally valid X-factor \citep[see recent review from][]{bolatto2013}. 
Consequently, investigating the dependence of the X-factor on differing 
physical conditions is clearly of great interest in establishing practical 
guidelines for its application. Combined CO-extinction studies of 
nearby clouds can therefore provide potentially important insights into this issue.

In this paper, we present observational results of molecular lines \cons(2-1), 
\thircons(2-1), and \ceions(2-1). The total gas mass (or N$_{\rm H_2}$) 
is traced by extinction derived from NIR observations, independent from the 
molecular lines. The area covered in our observation is mainly dense gas 
with \av $\ga$ 3 mag. With the NICEST technique \citep{2009A&A...493..735L}, 
we are able to trace extinction up to \av $\sim$ 33 mag, 
which is much deeper than achieved in a similar study of Orion by \citet{ripple13} 
who used the NICER technique \citep{2001A&A...377.1023L} to 
trace extinction depths up to \av $\sim$ 15 mag. 
We also observed the optically thinner 
\ceions(2-1) line, which provides insights into CO chemistry 
\citep[e.g.,][]{1994ApJ...429..694L,Shimajiri14} that cannot be readily 
derived using only \co and \thirco observations.

In the following, we describe our observations and data reduction in 
Section \ref{sec:obs}. We present the results and analysis in 
Section \ref{sec:res}. Discussions and conclusions will be presented in 
Section \ref{sec:discus} and Section \ref{sec:con}, respectively.

\section{Observations and Data Reduction}\label{sec:obs}

\subsection{Near-Infrared Wide-Field Mapping Survey}\label{subsec:nicest}

\subsubsection{Telescope and Observations \label{s:obs:ss:caha}}

\begin{deluxetable*}{llccccl}
\tabletypesize{\scriptsize}
\tablecolumns{7}
\tablewidth{0pc}
\tablecaption{Near-Infrared Observations of L1482 Fields\label{tab:obs}} 
\tablehead{\colhead{Field ID} &\colhead{Date Obs.} &\multicolumn{2}{c}{Center Coords.} &\colhead{Filter} &\colhead{Seeing} &\colhead{LF Peak\tablenotemark{a}}\\
\colhead{} &\colhead{} &\multicolumn{2}{c}{J2000} &\colhead{} &\colhead{[$(\arcsec)$]} &\colhead{[mag]}\\
\cline{1-7}
\multicolumn{7}{c}{CALAR ALTO 3.5 m OMEGA 2000 OBSERVATIONS}
}
\startdata
 CN-01    &   2009-12-24   &  04:30:58.81 & +34:51:57.5 &	$J$   &  1.49  &  19.00  \\
 CN-01    &   2009-12-24   &  04:30:58.81 & +34:51:57.5 &	$H$   &  1.52  &  19.75  \\
 CN-01    &   2009-12-24   &  04:30:58.81 & +34:51:57.5 &	$Ks$  &  1.39  &  19.25  \\
 CN-02    &   2009-12-08   &  04:30:38.55 & +35:04:20.4 &	$J$   &  1.36  &  19.50  \\
 CN-02    &   2009-10-09   &  04:30:38.55 & +35:04:20.4 &	$H$   &  1.59  &  20.75  \\
 CN-02    &   2009-10-09   &  04:30:38.55 & +35:04:20.4 &	$Ks$  &  1.47  &  19.75  \\
 CN-03    &   2012-12-23   &  04:30:07.84 & +35:16:06.6 &	$J$   &  1.11  &  22.25  \\
 CN-03    &   2012-12-23   &  04:30:07.84 & +35:16:06.6 &	$H$   &  1.06  &  20.25  \\
 CN-03    &   2012-12-23   &  04:30:07.84 & +35:16:06.6 &	$Ks$  &  1.12  &  19.25  \\
 CN-04    &   2009-10-07   &  04:30:44.04 & +35:27:49.5 &	$J$   &  1.05  &  21.25  \\
 CN-04    &   2009-10-07   &  04:30:44.04 & +35:27:49.5 &	$H$   &  0.96  &  19.50  \\
 CN-04    &   2009-10-09   &  04:30:44.04 & +35:27:49.5 &	$Ks$  &  1.05  &  21.50  \\
 CN-05    &   2009-10-07   &  04:30:44.04 & +35:41:24.1 &	$J$   &  0.91  &  21.25  \\
 CN-05    &   2009-10-07   &  04:30:44.04 & +35:41:24.1 &	$H$   &  0.93  &  22.00  \\
 CN-05    &   2009-10-06   &  04:30:44.04 & +35:41:24.1 &	$Ks$  &  1.17  &  19.75  \\
 CN-06    &   2009-10-06   &  04:30:44.73 & +35:54:54.8 &	$J$   &  1.29  &  22.00  \\
 CN-06    &   2009-10-06   &  04:30:44.73 & +35:54:54.8 &	$H$   &  1.26  &  21.00  \\
 CN-06    &   2009-10-06   &  04:30:44.73 & +35:54:54.8 &	$Ks$  &  1.30  &  20.00  \\
 CN-07    &   2010-01-02   &  04:29:08.58 & +36:29:18.7 &	$J$   &  1.07  &  20.75  \\
 CN-07    &   2010-01-02   &  04:29:08.58 & +36:29:18.7 &	$H$   &  1.13  &  19.00  \\
 CN-07    &   2010-01-02   &  04:29:08.58 & +36:29:18.7 &	$Ks$  &  1.18  &  18.75  \\
 CN-08    &   2010-01-03   &  04:28:16.37 & +36:29:58.4 &	$J$   &  1.37  &  20.25  \\
 CN-08    &   2010-01-03   &  04:28:16.37 & +36:29:58.4 &	$H$   &  1.35  &  20.00  \\
 CN-08    &   2010-01-03   &  04:28:16.37 & +36:29:58.4 &	$Ks$  &  1.02  &  19.25

 \enddata
\tablenotetext{a}{Expresses turnover point of observed magnitude distribution} 

\end{deluxetable*}

Near-IR observations for this study were obtained at the 3.5 m telescope 
of the Centro Astron\'omico Hispano Alem\'an observatory (hereafter CAHA) 
at Calar Alto in Almer\'ia, Spain. Our observations were 
made with the OMEGA 2000 camera, which has a field of view 
of 15$\arcmin$. In this paper we report observations for eight fields 
that cover the L1482 region. These fields are part of a more comprehensive 
survey of the CMC with a total of 24 fields. 
Fields CN01 to CN08 (Table \ref{tab:obs}) were observed in the three main near-IR broadbands, 
$J$, $H$ and $K_s$, between October 2009 and January 2010, with 
acceptable weather and seeing, except for one field, CN03, which 
was repeated on December, 2012. All observations consisted of 20 
exposures of 60 seconds, coadded as 1200 second exposure mosaics. 
The fields observed with the CAHA 3.5 m telescope were selected to 
cover the main filament or ``spine" of L1482, including the region of the 
LkH$\alpha$ 101 cluster. The pixel scale of OMEGA 2000 at the 3.5 m 
telescope is 0.45 $\mbox{arcsec pix}^{-1}$, with excellent uniformity 
and negligible geometric distortion across the FOV.

A list of all fields observed and considered for final analysis can be 
consulted in table \ref{tab:obs}. The table lists the field identification, 
the center of field positions, observation date, filter, an estimate of the 
seeing based on the average FWHM of the stars detected in each field, 
and the peak values for the brightness distributions, which are a good 
measurement of the sensitivity limits achieved. Below, we describe the 
observations and the data reduction process, including the construction 
of the photometric catalogs used to make our dust extinction map.

\subsubsection{Near-Infrared Data Reduction and Calibration \label{s:obs:ss:red}}

The near-IR image data from OMEGA 2000 was reduced with pipelines 
that made use of standard IRAF procedures complemented
with stand alone routines. These pipelines are based on the 
FLAMINGOS near-IR reduction, photometry and astrometry
pipelines \citep{Roman-Zuniga:2006aa,Levine:2006ab}. 
The data process is essentially identical to the one described in
\citet{2010ApJ...725.2232R} and we refer the reader to that paper for details.

Both photometry and astrometry of OMEGA 2000 data products were 
calibrated relative to Two-Micron All Sky Survey (2MASS) lists
obtained from the All-Sky Point Source Catalog (PSC). 
The final photometry catalogs were merged with a combination of 
TOPCAT-STIL \citep{2005ASPC..347...29T} and IDL routines. 
In the case of adjacent fields, overlapping areas were treated with 
a routine designed to handle duplicate detections in such a way that 
we list preferentially a higher quality observation (e.g. a smaller 
photometric uncertainty) over a lower quality one.

The 2MASS data were also used to replace observations for saturated 
stars in all our frames. We used, in all cases, lists retrieved from the 
2MASS All-Sky Point Source Catalog (PSC) at the Infrared Processing 
and Analysis Center (IPAC). The final photometry catalogs, containing either 
J, H, and Ks, or H and Ks photometry were merged with a combination of 
TOPCAT-STIL \citep{2005ASPC..347...29T} and self-made IDL routines. 

\subsection{The CO Molecular-Line Survey}

\subsubsection{Telescope and Observations }

\begin{figure}[htb!]
\epsscale{1.2}
\plotone{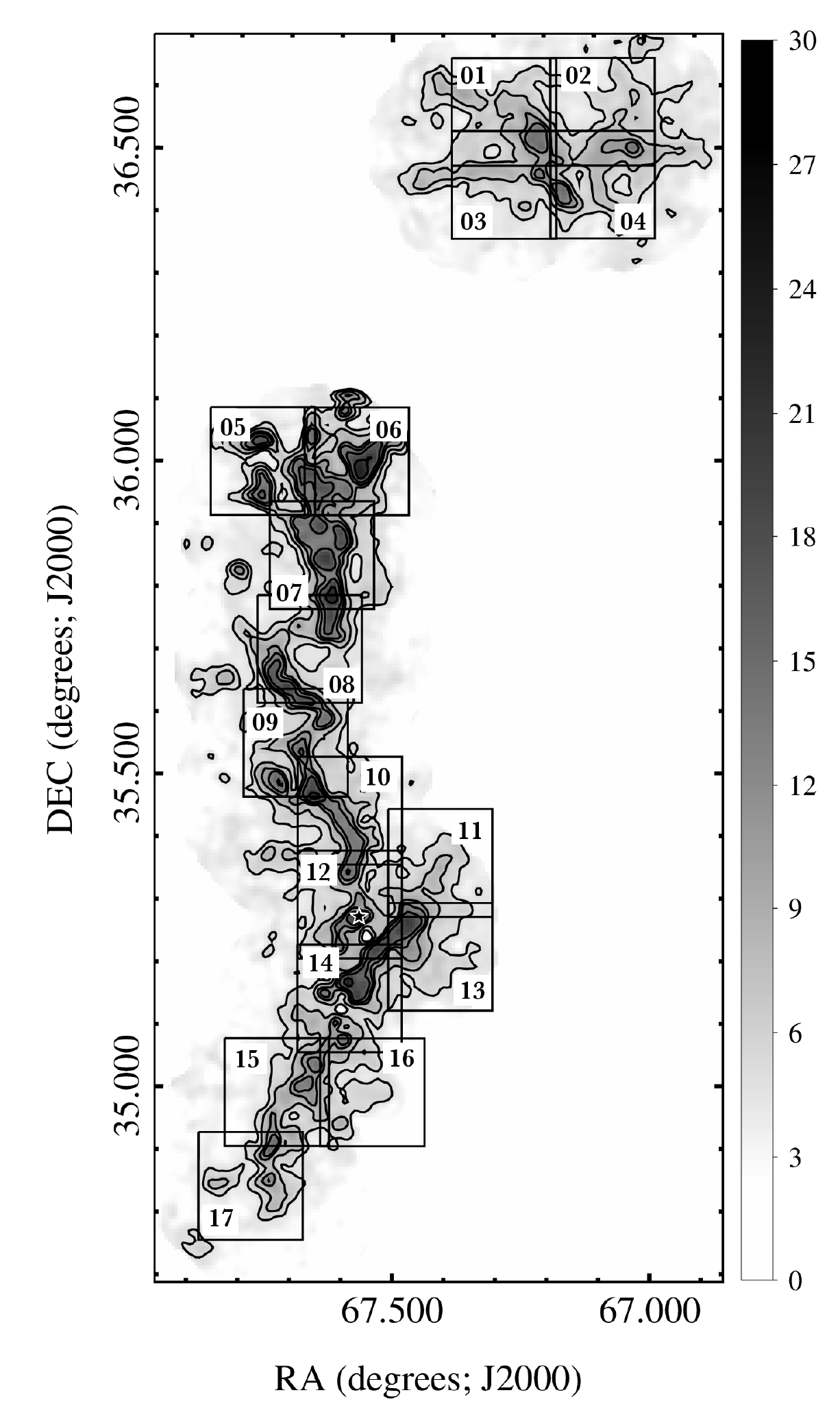}
\caption{
Contour map of visual extinction derived from our near-infrared survey with 
the NICEST method. Contours at intervals of \av= (5, 7, 10, 12, 15, 20, 25 and 30 mag). 
See descriptions in Section \ref{subsec:nicest}. The square boxes outline
the 17 tiles representing the areas surveyed in CO emission. 
LkH$\alpha$ 101 is marked by the star in tile 12.
\label{fig:obs}}
\end{figure}

A series of observations were carried out with the Heinrich Hertz 
Sub-millimeter Telescope (SMT) on Mount Graham, Arizona, from 
November 2011 to April 2012. The SMT is located at an elevation 
of 3200 meters. The receiver has prototype ALMA Band 6 
sideband separating mixers with two orthogonal polarizations. 
Typical single sideband system 
temperatures during the observing were around 200 K. 
The molecular lines \cons(2-1) (230.538 GHz), \thircons(2-1) 
(220.399 GHz), and \ceions(2-1) (219.560 GHz) were observed over 
a selection of 17 $10\arcmin\times10\arcmin$ square regions (hereafter ``tiles") 
along the south-eastern dense ridge of the CMC, roughly at 
RA=$\rm 04^h30^m00^s$, DEC=$35\arcdeg50\arcmin00\arcsec$ 
\citep[see Figure \ref{fig:obs}, also 
figure 1 of][]{lada09}. These tiles have \av $\ga$ 3 mag, 
and cover the higher extinction regions of the cloud as well as the 
LkH$\alpha$ 101 cluster (roughly centered in tile 12).

In each tile, we performed the observations using the ``on-the-fly (OTF)" mode. 
The scanning rate was $10\arcsec$ per second along the direction of Right Ascension (RA), 
with each row being separated by $10\arcsec$ in Declination (DEC). 
Each sampling point has a 0.4-second integration, corresponding to 
4$\arcsec$ spatial sampling in RA with a beam size around $35\arcsec$,
therefore the source is fully sampled. 
The total integration time for each tile is roughly 1 hour. We utilized 
dual side band mode and had two distinct sets of observations: 
 Obs1: \thircons(2-1) in lower side-band (LSB) 
and \cons(2-1) in upper side-band (USB); 
Obs2: \ceions(2-1) in LSB and \cons(2-1) in USB. 
In each set, two orthogonal linear polarizations were observed simultaneously. 
The total bandwidth we cover with the spectrometer is 128 channels of 0.25 MHz each or 32 MHz total, 
for each line (\cons, \thircons, and \ceions).  
The channel resolution is 0.25 MHz, set by the filter widths, 
and this corresponds to about 0.34 $\rm km~s^{-1}$ velocity resolution, 
depending on the line rest frequencies via the doppler formula.
We checked the pointing on the compact nearby CO source CRL618
(RA=$\rm 04^h43^m42^s.9$, DEC=$36\arcdeg08\arcmin14.7\arcsec$) 
at the beginning of every tile. Five-second integrations were performed 
on each point of a 5-point cross pattern centered on that source. 
The pointing error was typically $\la$ 5$\arcsec$. To calibrate the intensity, 
we observed the strong CO source W3OH 
(RA=$\rm02^h27^m59^s.2$, DEC=$61\arcdeg55\arcmin43.2\arcsec$) 
between every two tiles. 

\subsubsection{Millimeter-wave Data Reduction}

All data were first processed through the CLASS software package. 
We fitted linear baselines and subtracted them from the data. 
The data image cubes were ported into MIRIAD software format \citep{1995ASPC...77..433S} for further analysis and display. 
OTF data were interpolated onto regular grids with grid spacing of 10$\arcsec$ before being stored into the MIRIAD data set. 
Next the data were 
calibrated using W3OH. A scale factor is derived through 
the observation toward W3OH between every two tiles. The factor
is used to scale the antenna temperature $\rm T_A^*$ to the
main-beam brightness temperature \tmb
\citep[see][for details]{bieging10}.
For each molecular transition, the two 
polarizations were averaged to increase the SNR. 
Finally all 17 tiles were combined together to form a single map 
(weighted by their RMS noise at the small overlapped area). 
The original data cube, with $\sim$ 0.34 km s$^{-1}$ velocity channels 
(set by the 0.25 MHz filter bandwidth) were re-sampled onto a 0.15 km s$^{-1}$ 
spacing, by 3rd-order polynomial interpolation. This re-sampling puts all of the CO isotopologues 
on exactly the same velocity grid, so we can make ratio maps, for example.
We convolved all maps with Gaussian kernels 
(FWHM = 18.1$\arcsec$ for \co map, 
FWHM = 13.6$\arcsec$ for \thirco and \ceio maps)
to have a final map resolution (FWHM) of 38$\arcsec$ with a grid spacing of 19$\arcsec$
in order to match the resolution of 
and facilitate direct comparison with our dust extinction maps. 

We calculated the RMS noise for all emission-free channels in each 
map to be about 0.1 K for all three lines. 
We compared the integrated intensity maps of the two sets of 
\co maps by subtracting one from the other. We found 
a small spatial difference (typically $\la$ 10$\arcsec$ between each 
pair of corresponding tiles), and we shifted and re-gridded {\rm Obs1} 
\co and \thirco maps to {\rm Obs2}. We averaged the two 
sets of \co maps weighted by their RMS noise to
lower the noise in the averaged \co map to 0.07 K (RMS).

\section{Results and Analysis}\label{sec:res}

\subsection{The NICEST Dust Extinction Map \label{s:maps}}\label{subsec:extmap}

Figure \ref{fig:obs} shows the deep dust extinction map of the southeast portion 
of the CMC derived from our observations. For this map Calar Alto data 
were supplemented with data from 2MASS in some outer portions of the 
surveyed area where Calar Alto data were not available. 
Also shown are the boundaries of the 17 tiles of the multi-line CO 
mapping survey. The extinction map was constructed with the optimized 
version of the Near Infrared Color Excess Revised (NICER) technique, 
known as NICEST \citep{2001A&A...377.1023L,2009A&A...493..735L}. 
The NICER/NICEST technique allows us to measure dust extinction from 
the infrared excess in the colors of background stars caused by dust absorption. 
The excess is derived by assuming as intrinsic colors, the colors of stars  
 detected in a nearby off-cloud control field with negligible 
extinction (see Table 1). For the measurements of extinction, we avoid the 
use of sources with intrinsic color excess, such as candidate young 
stellar objects and dusty galaxies. NICEST is essentially identical to 
NICER, except that the \av estimator is modified to account for 
small-scale inhomogeneities, due to a bias introduced by the decrease in the number 
of observable stars as extinction increases toward the denser areas 
of a molecular cloud. The bias is removed by modifying the estimator with 
an additional term that accounts for the expected decrement. 

Extinction measurements for individual sources are smoothed with 
Gaussian filter with a FWHM of 38$\arcsec$, and the maps are constructed 
with 19$\arcsec$ spatial sampling.  
At each (line-of-sight) position  in the map, each star falling within the beam 
is given a weight calculated from a Gaussian function, and the inverse of 
the variance squared, which is in turn also used for the bias-correction factor. 
A preliminary value of \av at the map position is calculated as the weighted 
median of all possible values within the beam. Then, we estimate the mean 
absolute deviation (m.a.d.) of these values to remove large deviates using 
m.a.d.-clipping. The remaining values are used by the NICEST estimator to 
calculate the final \av at the corresponding position. 

In the case of pixel positions where only 2MASS sources are available, 
the number of sources per beam is significantly smaller, and there are 
cases where there are not enough stars to estimate \avns; in those cases, 
our code increases the size of the beam by a factor of 2 over the nominal value. 
The nominal 38$\arcsec$ FWHM of our maps represents an increase of resolution 
by a factor of 2.1 compared to the earlier study of \citet{lada09}. 
This resolution is comparable to the beam size of the CO maps.
The resulting noise of extinction map is 0.05 mag for \av $<$ 10 mag 
and 0.1 mag for \av $\geq$ 10 mag.

\subsection{CO Spectra and Integrated Intensity Maps}

\begin{figure*}[htb!]
\epsscale{1.1}
\plotone{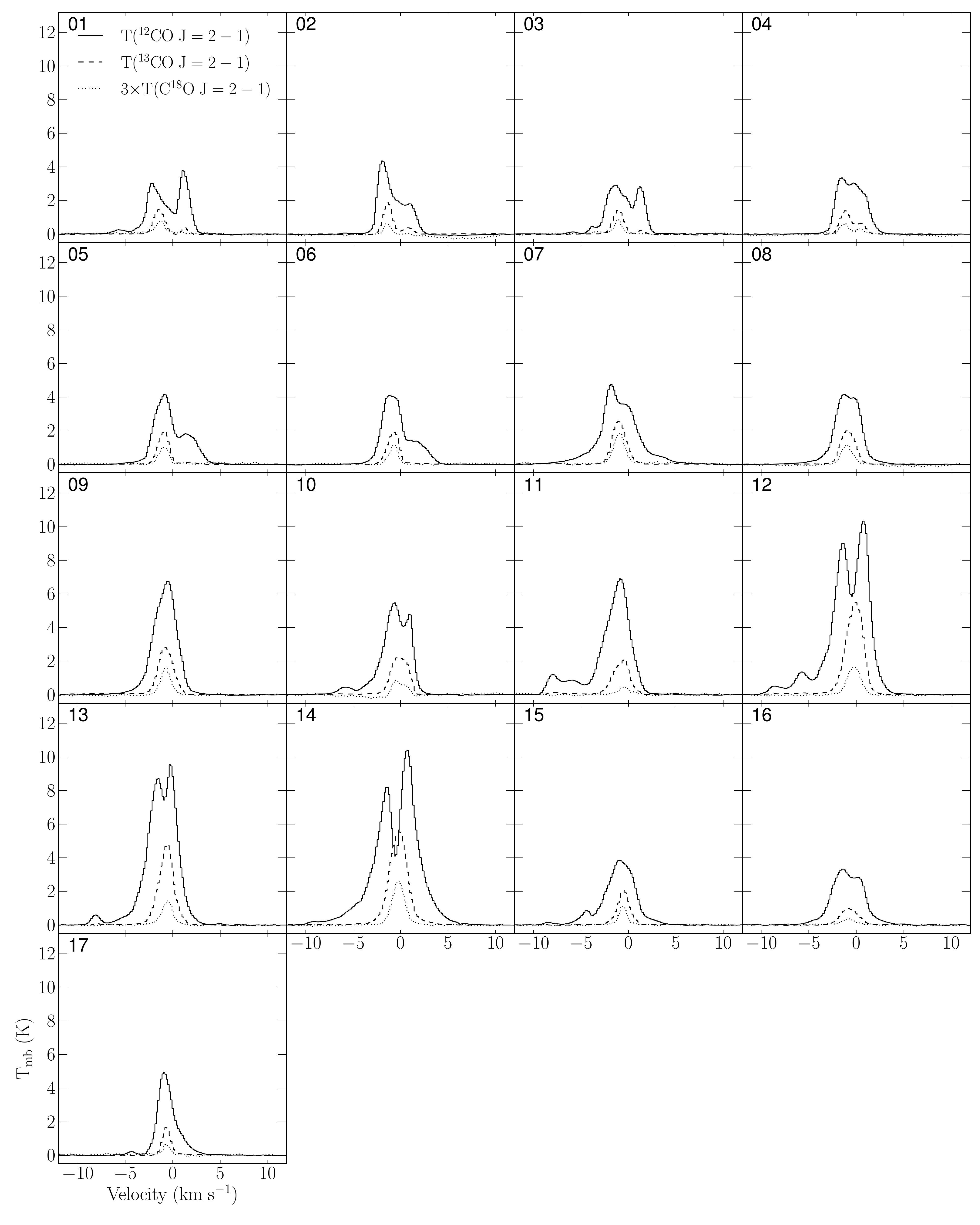}
\caption{
Averaged spectra for all three molecular lines in each 
$10\arcmin\times10\arcmin$ tile. Roughly 1000 spectra (from 1000 pixels) 
in each tile were averaged together. \ceions(2-1) 
spectra were enlarged by a factor of 3 to be shown together 
with the other two line spectra. 
\label{fig:spec}}
\end{figure*}

\begin{figure*}[htb!]
\epsscale{.38}
\hspace{1.5 cm}(a)\hspace{5.5 cm}(b)\hspace{5.5 cm}(c)\\
\plotone{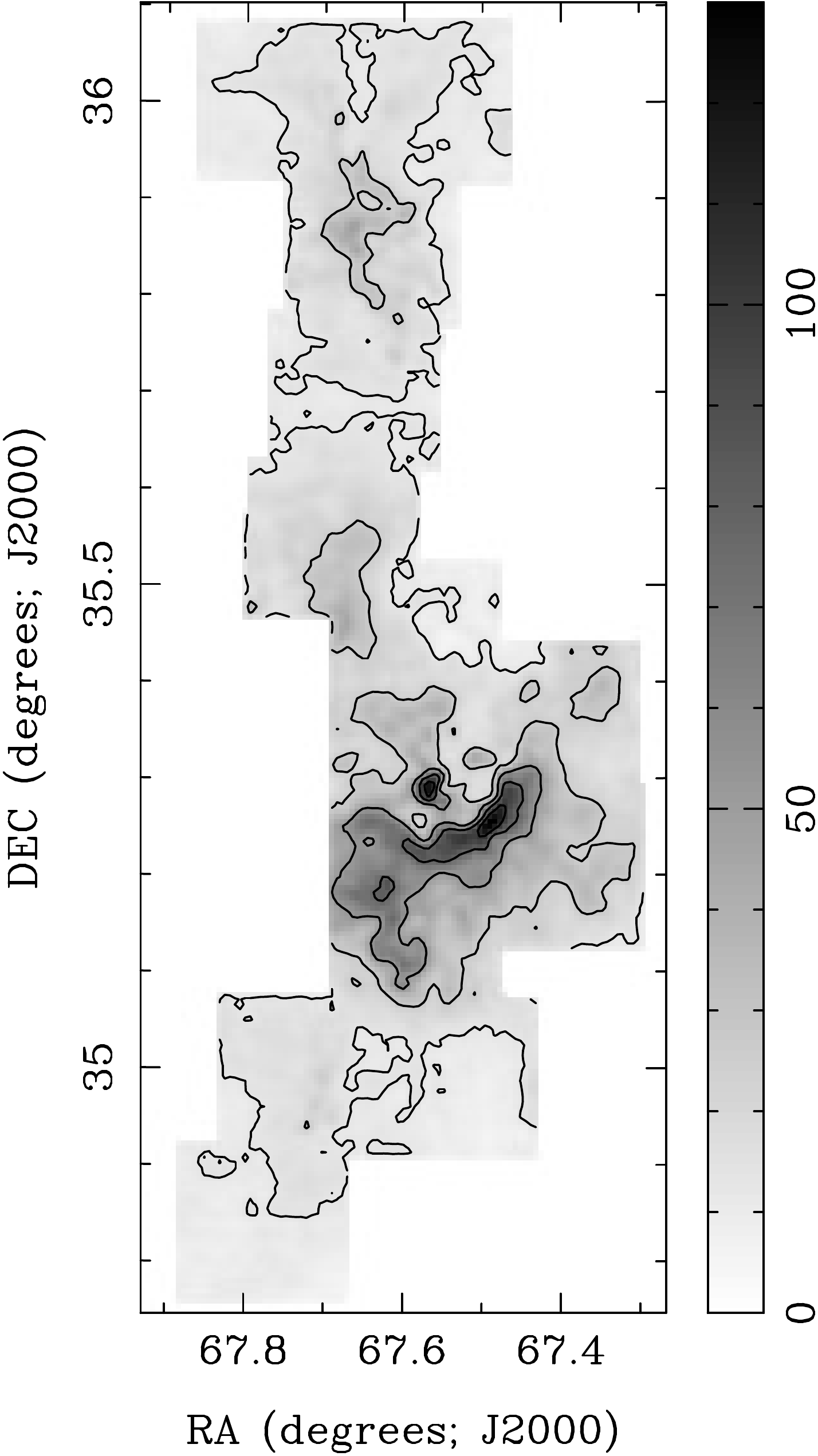}\plotone{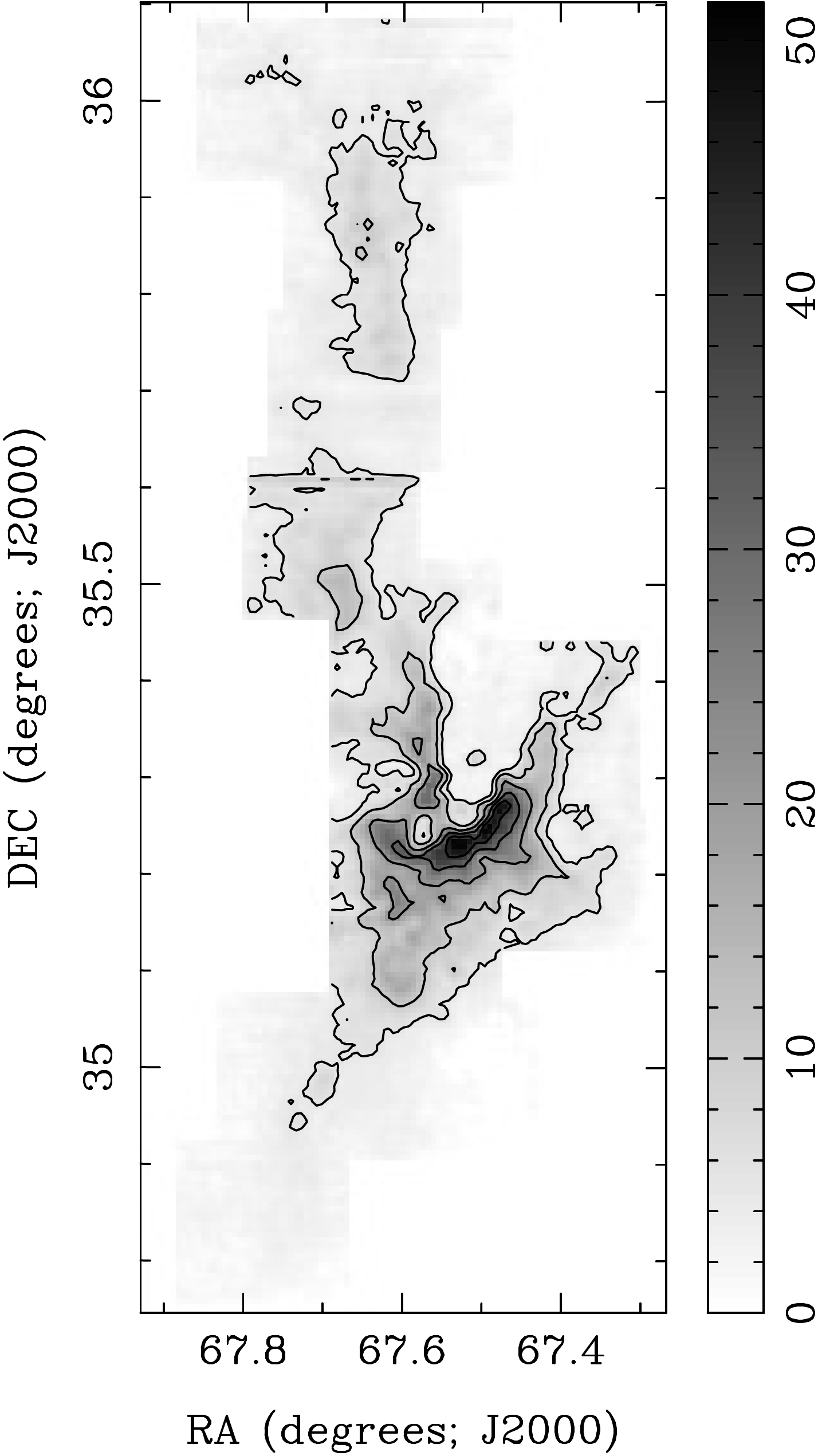}\plotone{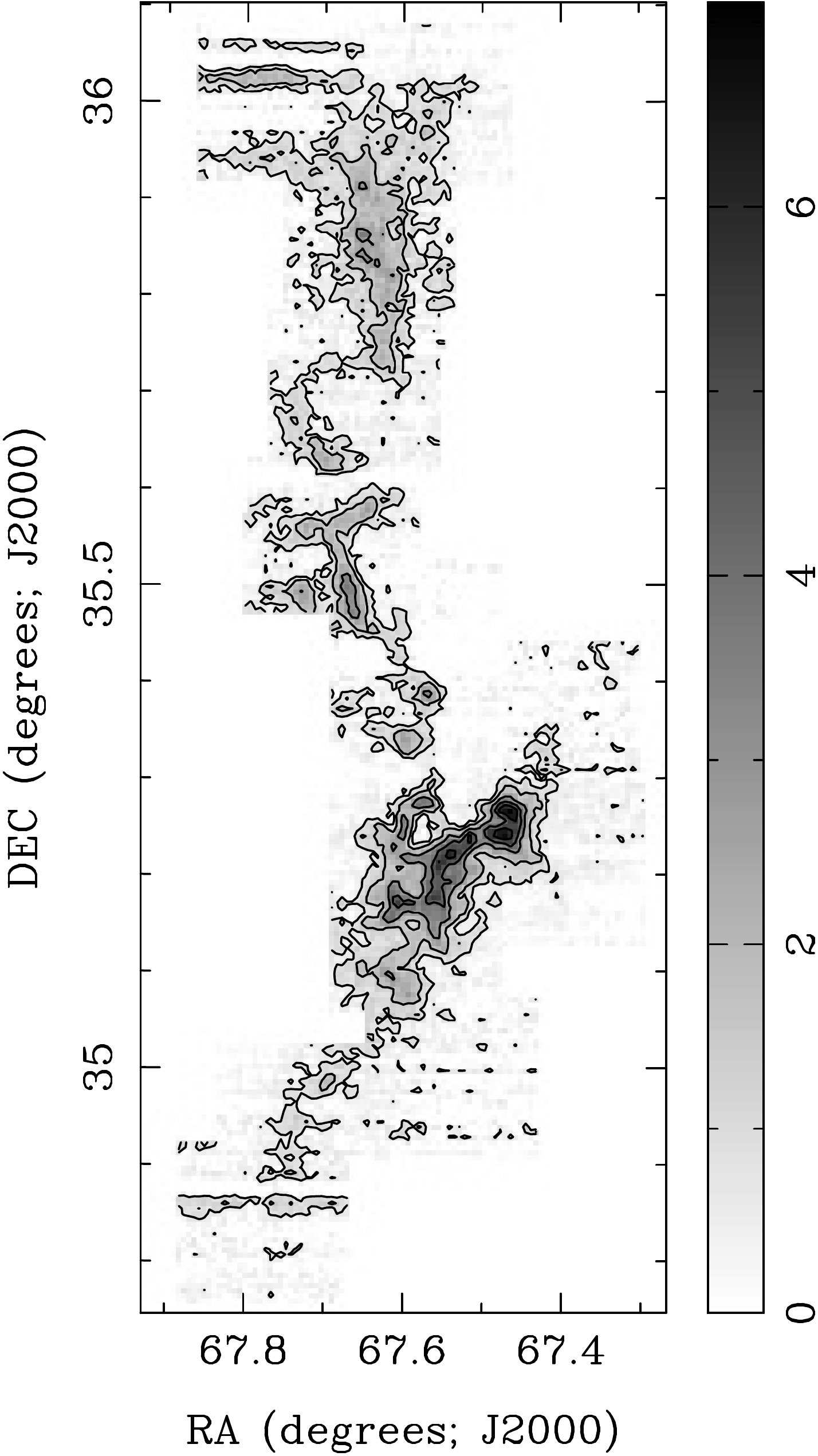}
\caption{
Integrated intensity maps of (a) \cons(2-1), 
(b) \thircons(2-1), (c) \ceions(2-1) of the 
Southeast region of the cloud shown in grey scale with contours overlaid for clarity. 
The integrated velocity ranges are -12 to 10 km s$^{-1}$, 
 -10 to 6 km s$^{-1}$, -4 to 3 km s$^{-1}$, respectively.
The maximum integrated intensities are 130.0, 51.5, 7.1 K km s$^{-1}$ for 
(a) \cons(2-1), (b) \thircons(2-1), (c) \ceions(2-1), respectively. 
The contour levels are 10\%, 20\%, 40\%, 60\%, 80\% of the maximum
of the respective maps.  
All the maps have been placed on  the same spatial 
grid with a resolution of 38$\arcsec$  sampled at 19$\arcsec$. 
See text for more details.
\label{fig:mom0}}
\end{figure*}

In order to provide an overview of the observations and the 
general quality of the data we first show in Figure \ref{fig:spec} 
the averaged CO spectra derived for each of the 
$10\arcmin\times10\arcmin$ tiles of our mapping grid. On this coarse scale 
(1.3 $\times$ 1.3 pc) the \co profiles exhibit  evidence for both 
self-absorption and multi-component structure. The presence 
of self-absorption indicates that the \co emission is 
quite optically thick. The profiles of the rarer isotopes appear 
to be less complex, mostly characterized by simple 
Gaussian-like shapes. Maps of the emission from 
\cons(2-1), \thircons(2-1), \ceions(2-1) integrated over velocity ranges, $-$12 to $+$10 km s$^{-1}$, 
 $-$10 to $+$6 km s$^{-1}$, $-$4 to $+$3 km s$^{-1}$, respectively, are shown in 
Figure \ref{fig:mom0} as grey scale maps with contours 
overlaid for clarity. The maps display some common 
morphological features. The strongest emission occurs 
in the mid-southern area (tiles 11-14, $\delta \approx  
35.25\degr$) and is in close proximity to the LkH$\alpha$ 
101 cluster (which lies within tiles 10-12, $\delta \approx 35.27\degr$). 
On close inspection of the maps, a sharp drop off in 
CO emission is observed along a warm filamentary structure 
southwest of the cluster.  This sharp drop in CO emission 
together with the enhanced CO emission within the warm 
filament likely indicates a physical interaction between the 
cluster stars and the filamentary ridge structure. A similar 
structure with an accompanying drop off is also present 
in the extinction map of Figure 1. Here we can see that the 
warm filamentary structure is a small section of the more 
extended high extinction ridge that forms the backbone 
of the cloud in this region and which is the subject of this survey. 
In these more extended regions away from the cluster, 
the CO emission is comparatively quiescent. 

The RMS noise for the integrated intensity maps $\sigma_W$
 (in K km s$^{-1}$) was first estimated using
\begin{equation}\label{eq:rms}
\sigma_W = \sigma_c\times\frac{\Delta V}{\sqrt{N_{\rm channel}}},
\end{equation}
where $\sigma_c$ (in K) is the RMS noise per channel derived for emission-free velocities, 
$\Delta V$ is the velocity width over which the integrated intensity is calculated.
N$_{\rm channel}$ here is the number of real independent channels or filters 
(of bandwidth 0.25 MHz, $\sim$ 0.34 km s$^{-1}$) within $\Delta V$. 
The estimated $\sigma_W$ for \cons(2-1), \thircons(2-1), \ceions(2-1) integrated 
intensity maps are 0.16, 0.22, 0.16 $\rm K~km~s^{-1}$, respectively. 
We also calculated the RMS for an integrated intensity map computed over 
the same number of emission free channels in the data cube. 
The $\sigma_W$ derived this way are 0.11, 0.24, 0.24 $\rm K~km~s^{-1}$, 
respectively.  To be conservative we adopt the larger value for $\sigma_W$, 
i.e., 0.16, 0.24, 0.24 $\rm K~km~s^{-1}$, for \cons(2-1), \thircons(2-1), \ceions(2-1), 
respectively (hereafter $\sigma_{W,12}$, $\sigma_{W,13}$, $\sigma_{W,18}$). 
The \co and \thirco lines are very strong and in our subsequent analysis we only consider 
integrated intensities detected above 5$\sigma_W$. With \ceions, we
only consider detections above 2$\sigma_W$.
We also note some striping apparent in tiles 5, 16 and 17 of 
the \ceio map. These are likely artifacts due to system 
temperature variations in some of the OTF scans in those regions of the cloud.
Prior to the subsequent analysis of these data reported below 
this map was ``cleaned" so that pixels containing the stripes 
were removed from the \ceio database\footnote{
In all cases, the stripes show up in only one polarization. We masked them before
combining the two polarization maps. The striping area would therefore have
higher noise (by a factor of 1.4). }

\subsection{CO excitation and column densities of \thirco and \ceio}\label{subsec:NAv}

In this paper we adopt the standard LTE analysis to 
determine excitation temperatures and column densities. 
\noindent
The main beam brightness temperature, \tmbns, is given by :  
\begin{equation}\label{eq:radt} 
T_{mb} =[ J_{\nu}(T_{ex}) - J_{\nu}(T_{bg})] (1 - e^{-\tau_\nu})
\end{equation}
\noindent
where $J_{\nu}(T) = {h\nu/ k \over (e^{h\nu/kT} -1)}$ is the Rayleigh-Jeans 
equivalent temperature of a blackbody of physical temperature T, 
\tex the excitation temperature of the line under consideration, 
\tbg the temperature of the background (in this case the CMB), 
and $\tau_\nu$  the optical depth of the transition.
The excitation temperature \tex of a given line 
can be derived from the peak temperature of the line, 
provided the line is optically thick.  Typically \co 
emission is optically thick in GMCs. In the CMC 
a large opacity for the \cons(2-1) line is confirmed by 
the \cons/\thirco integrated intensity ratio. 
The probability density distribution for this ratio for all pixels is 
found to peak near a value of 3.5, considerably smaller 
than the optically thin limit of $\approx$ 40-70 \citep[see, e.g.,][]{1993ApJ...408..539L}. 
Less than 10\% of the pixels are characterized by ratios 
exceeding 10. This is consistent with the fact that the 
surveyed area was dominated by relatively high 
extinction (dust column density). For $\tau \gg 1$, 
\tex for the \co line is calculated from 
the following equation (by solving Eq.\ref{eq:radt}):
\begin{equation}\label{eq:tex}
T_{ex}=\frac{h\nu/k}{ln(1+\frac{h\nu/k}{T_{\rm mb,p}+J_{\nu}(T_{bg})})}
\end{equation}
where T$_{\rm mb,p}$ is the peak brightness temperature in the \co line. 
At this frequency, $J_{\nu}(T_{bg})$ = 0.19 K and is generally much less 
than the peak line temperature (T$_{\rm mb,p}$). 
The typical error\footnote{This value encloses 90\% of pixel errors.} 
in \tex propagated from the 3 sigma uncertainty in T$_{\rm mb,p}$ is $\sim$ 0.22 K.
The maximum percentage error in \tex is below 5\%.
Note that in some regions, e.g., tiles 12 and 14, the \co line shows self-absorption,
this would result in underestimation of T$_{\rm mb,p}$ and thus \texns.
For T$_{\rm mb,p}$ between 6 K and 50 K, a change of this quantity by 1 K 
produces a corresponding change of \tex by 1 K.

\begin{figure}[htb!]
\epsscale{1.1}
\plotone{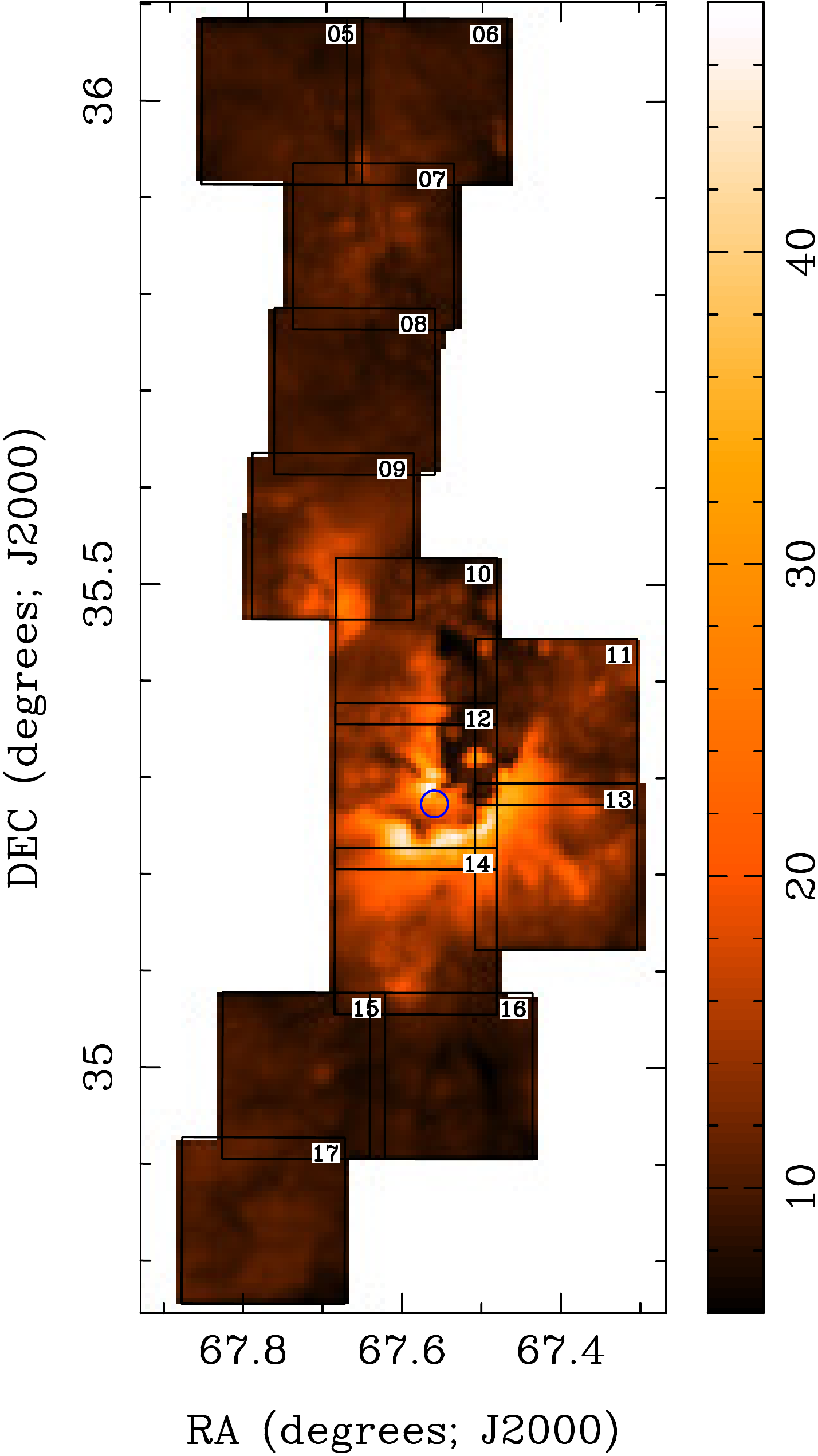}
\caption{
\tex map of the Southeast region of the cloud derived from CO(2-1) map 
using Equation \ref{eq:tex}. The boxes show the observed tiles 5-17. 
The blue open circle shows the position of the LkH$\alpha$ 101 
cluster (from SIMBAD).
\label{fig:texmap}}
\end{figure}

Figure \ref{fig:texmap} shows a map of the spatial 
distribution of the \co excitation temperatures 
in the surveyed region. Over the vast majority of the 
cloud, \tex is found to be in a narrow range around 10 K 
as expected for gas heated by cosmic rays and 
cooled by molecular lines \citep{1978ApJ...222..881G,2007ARA&A..45..339B}.  
Higher temperatures are found in the southern part of 
the region. The highest excitation temperatures 
approach 40 K and are found within the bright filamentary 
ridge southwest of the 
LkH$\alpha$ 101 cluster (see figure) suggesting that the 
cluster is providing  additional heating of the gas in that area.

The optical depths of \thircons(2-1) ($\tau_{13}$) 
and \ceions(2-1) ($\tau_{18}$) were calculated assuming LTE from:  
\begin{equation}\label{eq:tau}
\tau =-ln[1-\frac {T_{mb}} {h\nu /k}\frac {1} {(e^{\frac{h\nu}{kT_{ex}}}-1)^{-1}-(e^{\frac{h\nu}{kT_{bg}}}-1)^{-1}}].
\end{equation}
Here \tex for both transitions  was taken to 
be that derived for \cons(2-1). 
The probability density distribution of opacities for \thirco peaks 
around 0.5 with the majority (87\%) of pixels appearing to be 
relatively optically thin (i.e., $\tau$ $<$ 1). 
For \ceio the distribution peaks around 0.2, 
significantly less than that for \thirco (as expected), 
and with $\tau$ $<$ 1 for essentially all pixels in the map.

We followed \citet{pineda10} in calculating both 
the \thirco and \ceio column density using 
Eq.\ref{eq:ncol} for J=2-1 transition\footnote{We noticed 
the typo in equation (18) in \citet{pineda10}, as pointed 
out by \citet{ripple13}}.  The rotational constant B$_0$ 
and the Einstein coefficient A$_{2\rightarrow1}$ are taken 
from an online database\footnote{\url{http://home.strw.leidenuniv.nl/~moldata/CO.html}}. 
The partition function Q is estimated by Eq.\ref{eq:part}, 
which is accurate to 10 percent for T $>$ 5 K (note that in 
our data, \tex $>$ 5 K everywhere, as shown in 
Figure \ref{fig:texmap}). 

\begin{equation}\label{eq:ncol}
\begin{split}
N=\frac{8\pi k\nu^2}{hc^3A_{2\rightarrow1}}&\frac{e^{\frac{h\nu}{kT_{bg}}}-1}{e^{\frac{h\nu}{kT_{bg}}}-e^{\frac{h\nu}{kT_{ex}}}}\frac{Q}{(2J+1)e^{\frac{-hB_0J(J+1)}{kT_{ex}}}}\\
&\frac{\int \tau(v){\rm d}v}{\int [1-e^{-\tau(v)}]{\rm d}v}\int T_{mb}{\rm d}v~(cm^{-2})\\
\end{split}
\end{equation}

\begin{equation}\label{eq:part}
Q \equiv \Sigma_J(2J+1)e^{\frac{-hB_0J(J+1)}{kT_{ex}}} \simeq \frac{kT_{ex}}{hB_0}+\frac{1}{3}
\end{equation}

We note that sub-thermal excitation for the optically thin
\thircons(2-1) and \ceions(2-1) lines might be expected to 
be a concern in low extinction regions of the cloud where 
number densities may be lower than the critical density for 
thermal (LTE) excitation ($\approx$ 3 $\times$ 10$^4$ cm$^{-3}$). 
This could lead to underestimates of the CO column densities in 
the outer regions. However, since the studied area here is 
dominated by relatively high extinction (\av $\gtrsim$ 3-4 magnitudes) 
we expect that the effects of any sub-thermal excitation are minimized. 
We can roughly estimate the magnitude of this effect using the RADEX, 
non-LTE, radiative transfer code \citep{2007A&A...468..627V}. 
We find that with $\rm n(H_2) = 5 \times 10^3$ cm$^{-3}$, 
T$_{\rm gas}$ $=$ 10 K, and W(\thircons) $\approx$ 2.5  K km s$^{-1}$,  
N$_{\rm RADEX}$ $\approx$ 2.0 N$_{\rm LTE}$.

\section{Discussion}\label{sec:discus}

\subsection{CO Column Densities and Visual Extinction}\label{subsec:overallcomp}

\subsubsection{\thirco and \ceio Abundances}\label{subsubsec:1318abun}

Since our measurements of CO column densities and 
dust extinction are set to the same angular resolution 
and spatial grid, we are in a position to directly 
compare the two data sets. We note here that the 
angular scale of a pixel is 19$\arcsec$, corresponding to a 
spatial scale of $\sim$ 0.04 pc ($\sim$ 8000 AU) at a 
distance of 450 pc \citep{lada09}. 

\begin{figure*}[htb!]
\epsscale{.55}
\plotone{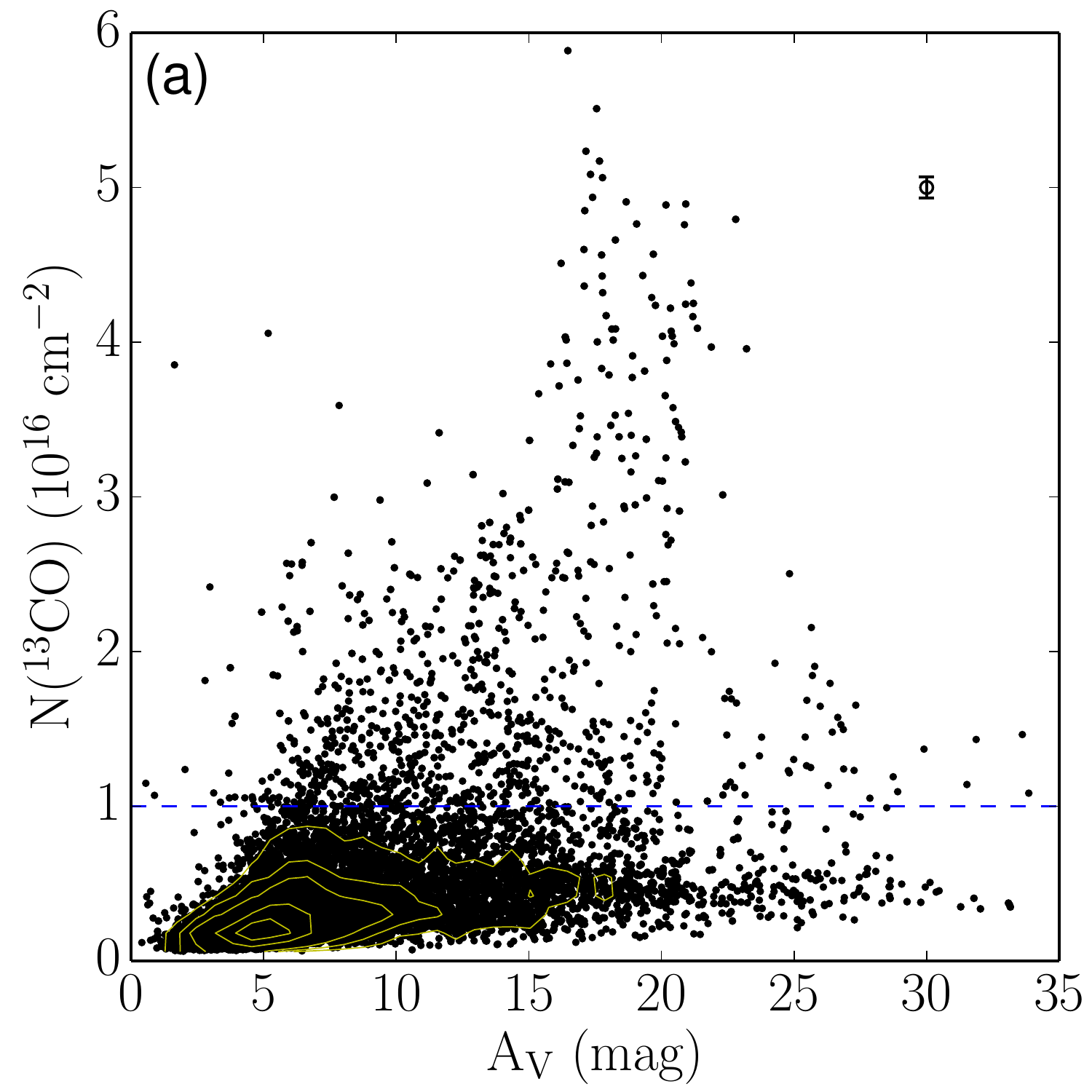}\plotone{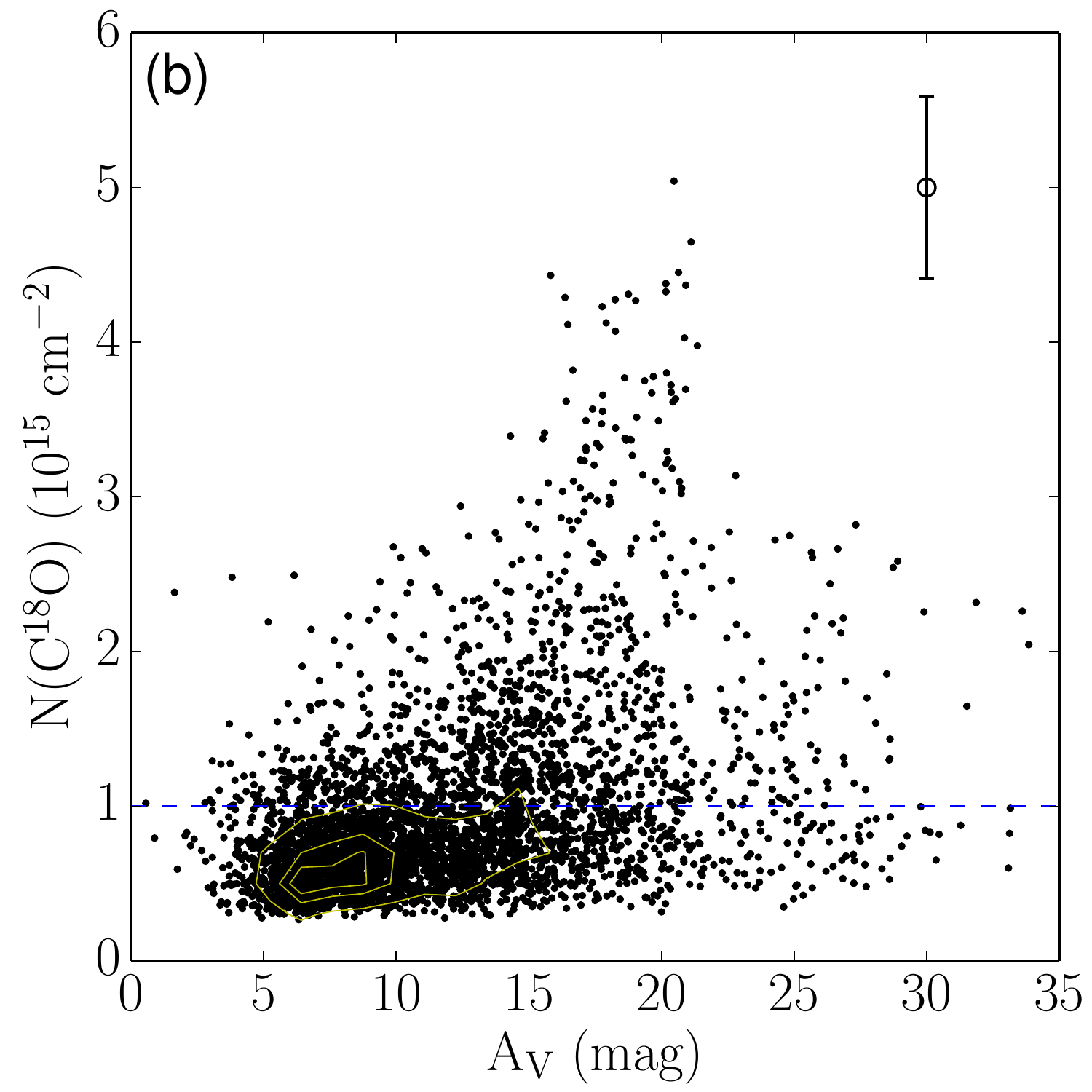}
\caption{
A pixel-by-pixel comparison between (a) \thircons, 
(b) \ceio column density and visual extinction 
\av for the entire surveyed area. Each point 
represents a 19$\arcsec$ pixel which corresponds to $\sim$ 0.04 
pc at a distance of 450 pc. 
A 5$\sigma_{W,13}$ cut is applied to W(\thircons) in \nthir calculation, 
while 2$\sigma_{W,13}$ cut is applied to W(\ceions) in \ne calculation
The blue dashed lines tentatively divide the points into two groups: 
an approximately constant abundance group above the line 
and a constant column density group below.
The size of the error bar encloses 90\% of the calculated 3 sigma errors
in column density (see text).
The error of \av is too small to be shown (Section \ref{subsec:extmap}).
See Section \ref{subsubsec:1318abun}.
\label{fig:ncol}}
\end{figure*}

Assuming a constant gas-to-dust mass ratio,
(N(H)/\av=1.88$\times$10$^{21}$ cm$^{-2}$ 
mag$^{-1}$) \citep{bohlin78,rachford2009,pineda08}, 
the CO abundance can be expressed as:  [CO] = N(CO)/N(H) = N(CO)/[\avns($\rm 1.88\times10^{21}~cm^{-2}~mag^{-1}$)].
For optically thin emission lines, such as \thirco and \ceions, 
we expect the gas column density to increase linearly 
with \avns, provided the cloud is characterized by a 
constant CO abundance \citep[e.g.,][]{1994ApJ...429..694L,1999ApJ...515..265A}. 
In Figure \ref{fig:ncol}  we plot the 
molecular column density vs. extinction at each pixel in 
the mapped region for \thirco (a) and \ceio (b), 
respectively.  Contrary to our simple expectations, 
each plot exhibits considerable scatter with only weak 
correlations between gas and dust column densities.
If the gas in the CMC were characterized by  a single, 
constant abundance, the relation between its column 
density and extinction would consist of a single straight 
line in each plot. At fixed \av there would be 
only small scatter in N(CO), which is clearly not the case.   
For instance, at \av $\sim$ 20 mag the ratio 
between the highest and lowest values of N(\thircons) 
and N(\ceions) is $\ga$ 10 for each isotope.  
Clearly, the two plots suggest strong spatial variations in 
abundances of these two species across the cloud.

On closer inspection of Figure \ref{fig:ncol}  one can almost 
make out two significant branches (correlations) of pixels 
in both plots. First, a large group of pixels in each plot 
(below the blue dashed line, i.e., N(\thircons) $\la$ 
1.0$\times$10$^{16}$ cm$^{-2}$, N(\ceions) $\la$ 
1.0$\times$10$^{15}$ cm$^{-2}$)
display an almost constant, low column density across 
a wide range of \avns. This flat relation between 
the two quantities indicates a trend of {\it decreasing} 
molecular abundance ([\thircons] and [\ceions]) 
from low to high \av pixels. Second, another group 
(above the blue dashed line)
of pixels is characterized by gas column densities that 
appear to rise more or less linearly with extinction to  very 
high column density at \av $\sim$ 20 mag 
(1.0$\times$10$^{16}$ cm$^{-2}$ $\la$ N(\thircons) 
$\la$ 6.0$\times$10$^{16}$ cm$^{-2}$, 1.0$\times$10$^{15}$ 
cm$^{-2}$ $\la$ N(\ceions) $\la$ 6.0$\times$10$^{15}$ 
cm$^{-2}$), indicating a more or less uniform molecular 
abundance with depth into the cloud for this group of pixels. 
These branches likely mark two limiting extremes of spatially 
dependent variation in the chemical conditions within the CMC.

 Spatial variations in \thirco abundances have been 
 reported previously in a few other molecular clouds, 
 although in these clouds the variations appear less 
 pronounced than found here. For the Perseus molecular cloud, 
 \citet{pineda08} found generally tighter correlations of \thirco column 
 density with extinction than found here, although they 
 explored the correlation at much lower extinctions (i.e., 
 \av $\la$ 10 mag) than in this study. Nonetheless, over 
 this lower extinction range they did find measurable 
 variations (at the $\sim$ 40\% level) in the abundances 
 with position across the cloud. For the Orion A molecular cloud, 
 \citet{ripple13}  measured abundances at both low and high extinction 
 and found more significant positional variations in the 
 \thirco abundance across the cloud than observed 
 in Perseus \citep{pineda08}. At \av $\sim$ 20 mag, N(\thircons) 
 in Orion A was found to vary by about a factor of 4 between 
 various locations in the cloud. Further, \citet{ripple13} were 
 able to associate these positional variations in abundance 
 with positional variations in the physical conditions within 
 the cloud. At the lowest extinctions (\av $\la$ 3 mag), 
 where there is not enough \thirco to self-shield against 
 UV dissociation, they found extremely low \thirco 
 abundances with significant (factor of 8) spatial variations. 
 At intermediate extinctions (3 $\la$ \av $\la$ 10 mag), where 
 self-shielding is considerably more effective, higher 
 abundances were found with modest (factor of 2) 
 spatial variations. At the highest extinctions they observed 
 the highest abundances but also the largest spatial 
 variations in abundances across the cloud.  This latter 
 regime included cold portions of the cloud where CO 
 depletion depressed the abundances as well as hot 
 regions heated by young stars where CO desorption 
 produced enhanced CO abundances.

\subsubsection{Mapping Spatial Variations in CO Abundances}\label{subsec:tilewav}

The molecular abundance is regulated by the physical 
and chemical conditions in the GMC. For instance, 
the gas-phase molecules can condense onto dust 
grains to form ice mantles, and the molecules on grain 
surfaces can be returned to the gas via thermal evaporation
\citep[e.g.,][]{1993prpl.conf..163V,1999ApJ...523L.165C}. 
These processes are tightly linked to the local temperature 
(especially the dust temperature), and are responsible 
for setting the gas-phase molecular abundance. 
At low extinctions where column densities of dust and 
gas are low, abundances are very sensitive to FUV 
radiation which can drive volatile chemistry via processes 
of fractionation and selective photodissociation 
\citep{1994ApJ...429..694L,2013A&A...550A..56R}. 
The large scatter in Figure \ref{fig:ncol} 
could plausibly be caused by strong variation of such 
physical/chemical conditions in different sub regions of 
the surveyed area of the cloud. If this is the case, 
we would expect to observe a tighter correlation 
between column density  and \av if we restricted 
the plots to cover smaller spatial areas within the surveyed 
region, since the physical conditions in such smaller regions 
would be expected to be more uniform. Indeed, 
\citet{ripple13} demonstrated the connection 
between such physical conditions and spatial CO 
abundance variations in the Orion cloud.

\begin{figure*}[htb!]
\epsscale{1.1}
\plotone{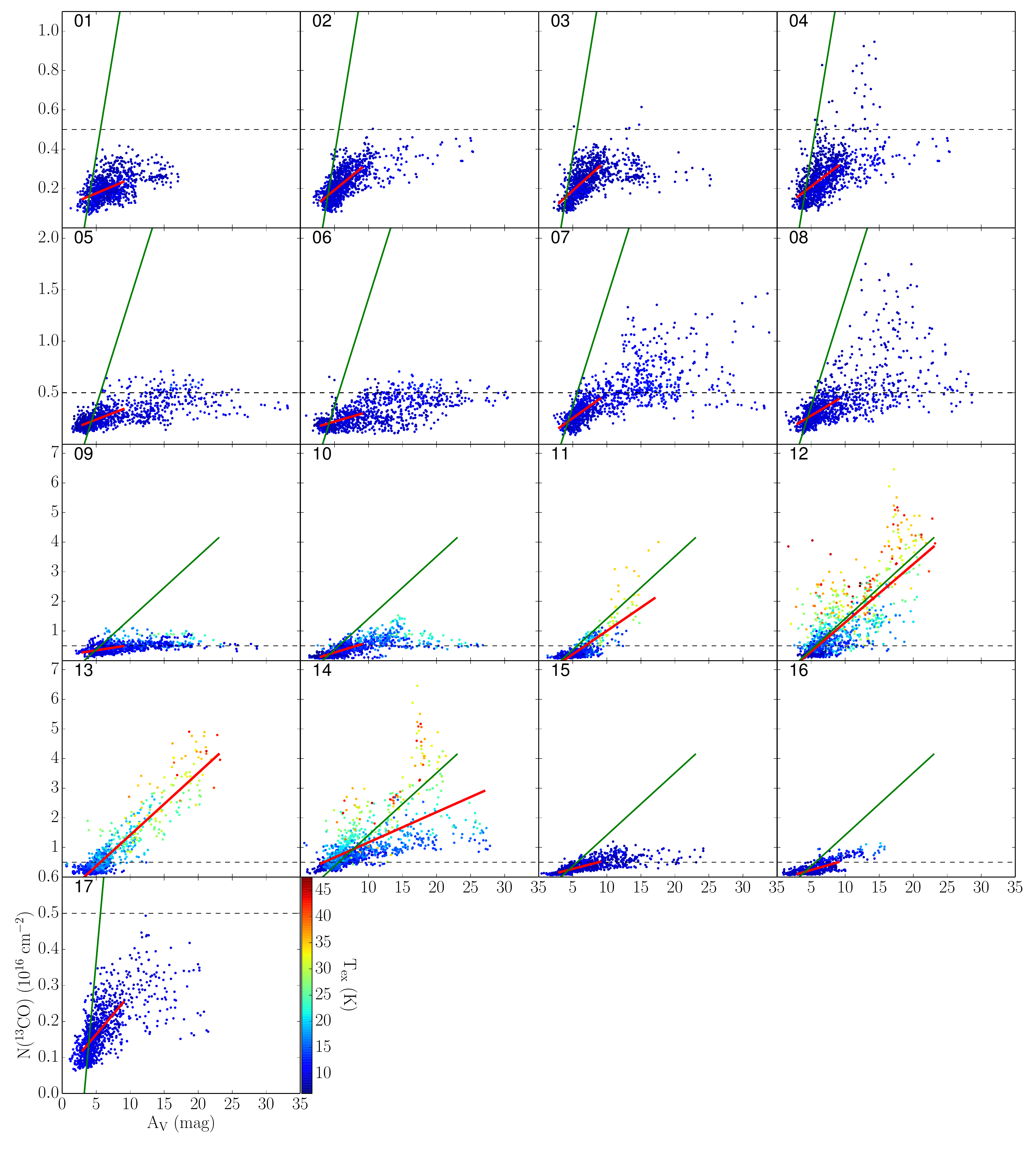}
\caption{
A pixel-by-pixel comparison between \thirco 
column density and \av within each tile. 
The scale of the column density axis varies between tile rows 
and a fiducial horizontal line representing a constant N(\thircons) 
of 0.5 $\times$ 10$^{16}$ cm$^{-2}$ is drawn for inter comparison. 
The individual points are color-coded by \texns. 
A 5$\sigma_{W,13}$ cut is applied to W(\thircons)
in \nthir calculation.
The red solid lines show linear regression in \av $>$ 3 mag
for tiles 11-14, and in 3 mag $<$ \av $<$ 10 mag for the rest.
As a reference, the fitting result for tile 13 is shown as green lines
in every tile. 
The magnitude of the error is the same as in Figure \ref{fig:ncol}(a).
}
\label{fig:ncol13} 
\end{figure*}

\begin{deluxetable}{cccc}
\tabletypesize{\scriptsize}
\tablecolumns{4}
\tablewidth{0pc}
\tablecaption{\thirco Survey Tile Properties\label{tab:tp}}
\tablehead{
\colhead{Tile} & \colhead{$\rm \overline{T}_{ex}$} & \colhead{slope\tablenotemark{a}}&\colhead{r-value\tablenotemark{a}}
\\
\colhead{}  & \colhead{(K)}   & \colhead{(10$^{16}$ cm$^{-2}$ mag$^{-1}$)} & \colhead{}
}
\startdata
01 &9.21&0.014&0.42\\
02 &9.31&0.027&0.65\\
03 &8.30&0.031&0.64\\
04 &8.52&0.025&0.50\\
05 &9.14&0.024&0.57\\
06 &9.23&0.018&0.43\\
07 &9.99&0.046&0.64\\
08 &9.19&0.037&0.55\\
09 &12.3&0.034&0.37\\
10 &12.3&0.066&0.67\\
11 &13.1&0.148&0.80\\
12 &19.6&0.193&0.76\\
13 &16.5&0.204&0.91\\
14 &16.9&0.102&0.62\\
15 &9.07&0.059&0.75\\
16 &8.65&0.054&0.64\\
17 &9.66&0.022&0.64
\enddata

\tablenotetext{a}{~Linear regression results from Figure \ref{fig:ncol13}. 
In tiles 11-14, the fitting is performed over the entire \av range. 
In the rest of tiles, the fitting is up to \av = 10 mag. The r-value (Pearson coefficient)
is an estimation of correlation coefficient.}

\end{deluxetable}

\begin{figure*}[htb!]
\epsscale{1.1}
\plotone{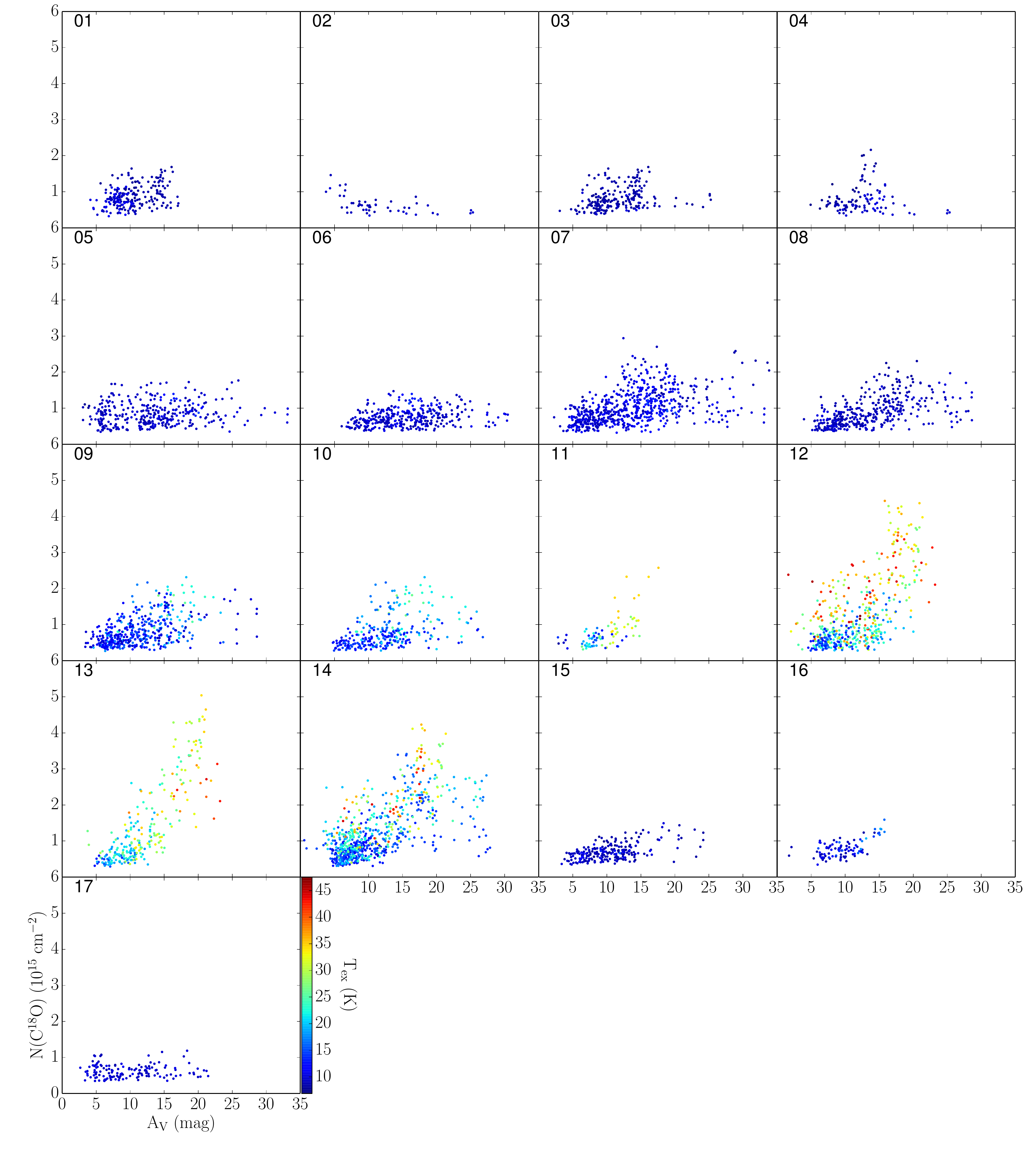}
\caption{
A pixel-by-pixel comparison between \ceio  
column density and \av within each tile. 
A 2$\sigma_{W,18}$ cut is applied to the integrated intensity
in \ne calculation.
The magnitude of the error is the same as in Figure \ref{fig:ncol}(b).
The axes have the same horizontal and vertical scales for all tiles, 
otherwise the same as Figure \ref{fig:ncol13}.
}
 \label{fig:ncol18}
\end{figure*}

In order to examine the distributions of N(\thircons) 
and N(\ceions) versus \av  on smaller 
spatial scales, we constructed these relations for 
each individually observed $10\arcmin\times10\arcmin$
mapping tile (c.f. Figures 1 and 2).  In Figures \ref{fig:ncol13} 
and \ref{fig:ncol18} we plot the results for \thirco and 
\ceions, respectively.  In the individual plots contained 
within these figures we indeed see a significant reduction 
in scatter compared to that characterizing Figure \ref{fig:ncol}. 
More importantly, there is very strong spatial variation in 
the relation between extinction and CO column density 
across the surveyed region of the CMC. 
This is evident in Figure 6 where we plot linear fits
to the data. The results of the fits are shown in Table 2 where
the slopes vary by nearly an order of magnitude across the 
cloud. This is also readily 
apparent in Figure \ref{fig:ncol18} where the axes on the 
plots have the same scaling and especially clear on close 
inspection of Figure \ref{fig:ncol13} where the scale of the 
vertical axis ranges by as much as an order of magnitude 
between the various tiles. These spatial variations in \nthir 
and \ne give rise to the large dispersion in these quantities 
in the merged relation constructed for the entire observed 
region as in Figure \ref{fig:ncol}. 
Despite the variation in the range of column densities at 
different locations across the cloud, the range in extinction 
spanned by the observations in all the plots is very nearly 
the same.
The large variation of column density of \thirco and \ceio with dust extinction
confirms that the abundances of the two 
isotopologues exhibit large  spatial variations. 
Consider Figure \ref{fig:ncol13}. 
The tiles can be placed into two groups, one containing 
tiles 1-10 and 15-17, and the other consisting of tiles 11-14. 
In the former group, \nthir rarely exceeds a 5-6 $\times$10$^{15}$ cm$^{-2}$, 
while in the latter group \nthir can reach values as high as 
3-5 $\times$ 10$^{16}$ cm$^{-2}$ over the same extinction range. 
The \ne plots in Figure \ref{fig:ncol18} exhibit similar behavior. 

\subsubsection{The Relation between Gas Temperature and CO Abundances}\label{subsubsec:texabun}

From comparison of Figures \ref{fig:ncol18} and \ref{fig:ncol13} 
with Figure \ref{fig:texmap} we see that the 
tiles (11-14) with the largest values of CO column density 
spatially coincide with the region of elevated \co excitation temperature. 
Since the \cons(2-1) line is generally so optically thick, 
and since our sampled region preferentially contains high 
column density (\av $\ga$ 3-5 mag) gas, 
sub-thermal excitation of \co is unlikely. Therefore 
the excitation temperature map traces 
the gas kinetic temperature and the distribution of the 
molecular abundances [\thircons] and [\ceions] 
are tightly related to local physical conditions, in particular 
the gas kinetic temperature. 
As a further test of the connection between column density and 
temperature we have color-coded the points in Figures \ref{fig:ncol13} 
and \ref{fig:ncol18} according to the gas temperature in each individual pixel.
Inspection of the figures shows that the pixels 
with the highest gas temperatures (\tex $\ga$ 15 K) are 
also the pixels with the largest CO column densities. 
Pixels with low gas temperatures (\tex $\la$ 15 K) 
rarely reach high CO column densities, independent of extinction. 
These correlations imply that the abundances of \thirco and \ceio 
are physically related to the gas temperature distribution.

\begin{figure*}[htb!]
\epsscale{.55}
\plotone{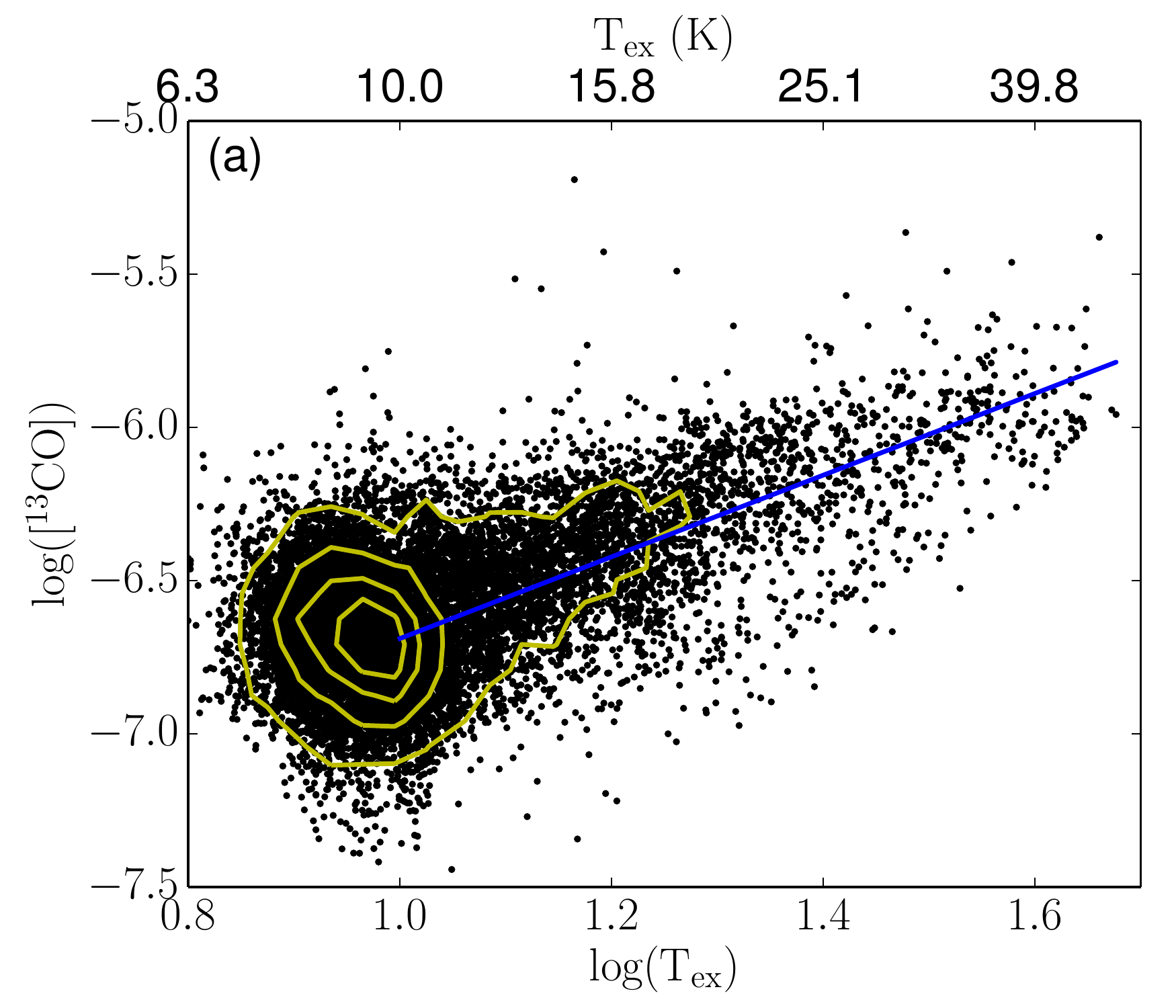}\plotone{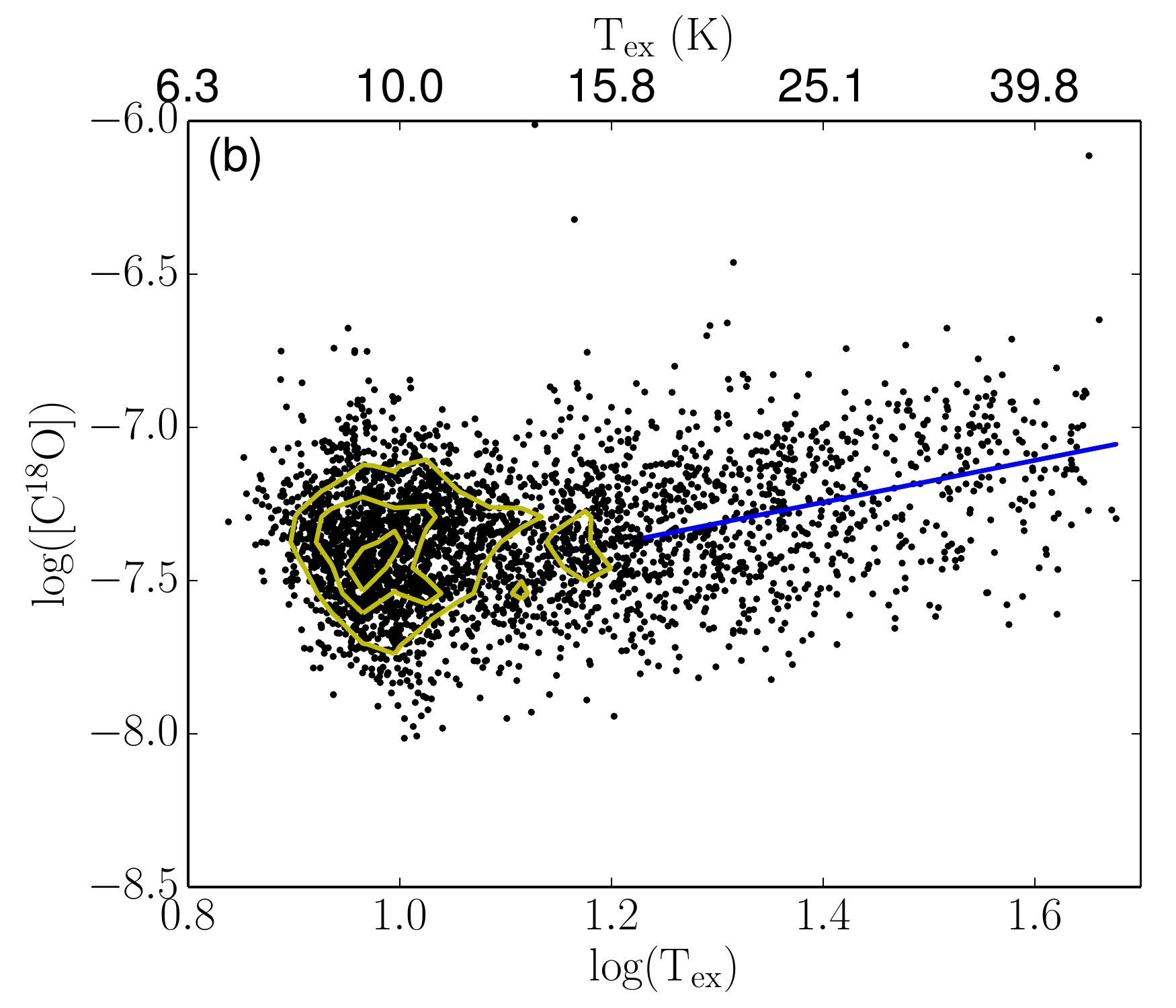}
\caption{
A pixel-by-pixel comparison between (a) \thircons, (b) \ceio abundance
and \tex for the entire surveyed area. Each point 
represents a 19$\arcsec$ pixel which corresponds to $\sim$ 0.04 pc 
at a distance of 450 pc. Abundances were calculated from pixels where
W(\thircons) $>$ 5$\sigma_{W,13}$ and W(\ceions) $>$ 3$\sigma_{W,18}$.
The yellow contours show the surface density of points.
The blue lines show the linear fitting within \tex $>$ 10 K for \thirco and \tex $>$ 17 K for \ceions.
See section \ref{subsubsec:texabun}.
\label{fig:abuntex}}
\end{figure*}

To examine more directly the relation between the 
molecular abundances and the gas temperature, 
we plot the log([\thircons]) - log(\texns) relation
and log([\ceions]) - log(\texns) relation in Figure \ref{fig:abuntex}.  
Both molecular abundances show dependence on \texns, although [\thircons] 
is apparently sensitive to \tex over a larger range of 
temperature than [\ceions].  
This confirms that [\thircons] and [\ceions] are both regulated
by gas temperature, but over different ranges of 
temperature in the extinction range
(3 mag $\la$ \av $\la$ 25 mag) we observed.
To quantify the relations, we empirically fit a power-law functions
to the unbinned data for both correlations. \tex is derived 
from the \co line, which has very good SNR. 
Since the noise of extinction map is relatively small, i.e., 0.05 mag for Av$<$10 mag 
and 0.1 mag for Av$\geq$10 mag, uncertainties in the abundances are
dominated by the noise in the molecular line data.
 
The fits were performed over different ranges of \tex for the two lines: 
\tex $>$ 10 K for \thirco and \tex $>$ 17 K for \ceions. 
The low temperature cutoffs were empirically chosen, by inspection,  
to correspond to the ranges where the relations appeared to be linear. 
However, it did not escape our attention that the low temperature 
cutoff for \ceio corresponds to the CO sublimation temperature of 
$\sim$ 17 K \citep{1988ApJ...334..771V}, and as discussed later this may not be coincidental.
The linear regression gave the following results:
[\thircons] $\propto$ T$_{\rm ex}^{1.33}$ with correlation coefficient = 0.66; 
[\ceions] $\propto$ T$_{\rm ex}^{0.69}$ with correlation coefficient = 0.34. 
We performed a $\chi^2$ test on the power-law model for [\thircons] - \tex and 
[\ceions] - \tex relations. The $\chi^2$ is calculated using:
\begin{equation}\label{eq:chi2}
\rm \chi^2 = \Sigma \frac{(O-E)^2}{\sigma_a^2}
\end{equation}
where O and E stand for observed and modeled [CO] (calculated from the
fitted power-law), respectively, and $\rm \sigma_a$ the local standard deviation
derived in a 2 K bin enclosing the data. At a significance level of 0.05, the fitted 
power-law models for [\thircons] - \tex and [\ceions] - \tex relations are acceptable,
and the $\chi^2$ per degree of freedom are 1.02 and 1.06 for [\thircons] and [\ceions], 
respectively, meaning the fit is reasonable.
The implications of these results will be discussed in the next two sections of the paper.

\subsubsection{Molecular Depletion and Desorption}\label{subsec:deple}

We argue here that the dependence of the \thirco 
and \ceio abundances on gas temperature shown 
in Figure \ref{fig:abuntex} is a result of the combined 
effects of gas depletion and desorption in the CMC. 
Theoretical considerations suggest that sticking of CO onto dust grains can remove the
molecule from gas-phase very efficiently within cold (T$<$ 17 K) 
and dense (n(H$_2$) $>$ 10$^4$  cm$^{-3}$) regions of molecular 
clouds \citep[see, e.g.,][]{1993prpl.conf..163V}.
Indeed, compelling observational evidence for reduction of [\ceions] 
has been reported  for numerous cold (T $<$ 10 K) and high extinction 
(\av $\ga$ 10 magnitudes) regions of molecular clouds 
\citep[e.g.,][]{1994ApJ...429..694L,1999ApJ...515..265A,1999ApJ...523L.165C,1999A&A...342..257K,2001ApJ...557..209B,2002ApJ...570L.101B}.
Both the radial stratification of the clouds \citep[e.g.,][]{1998ApJ...506..292A,2001Natur.409..159A,2001ApJ...557..209B}
and the presence of emission from high dipole molecules such as 
NH$_3$ from such regions indicate that they likely correspond to 
sufficiently high gas volume densities ($>$10$^4$ cm$^{-3}$) to 
promote CO condensation onto grain surfaces.

\begin{figure*}[htb!]
\epsscale{.55}
\plotone{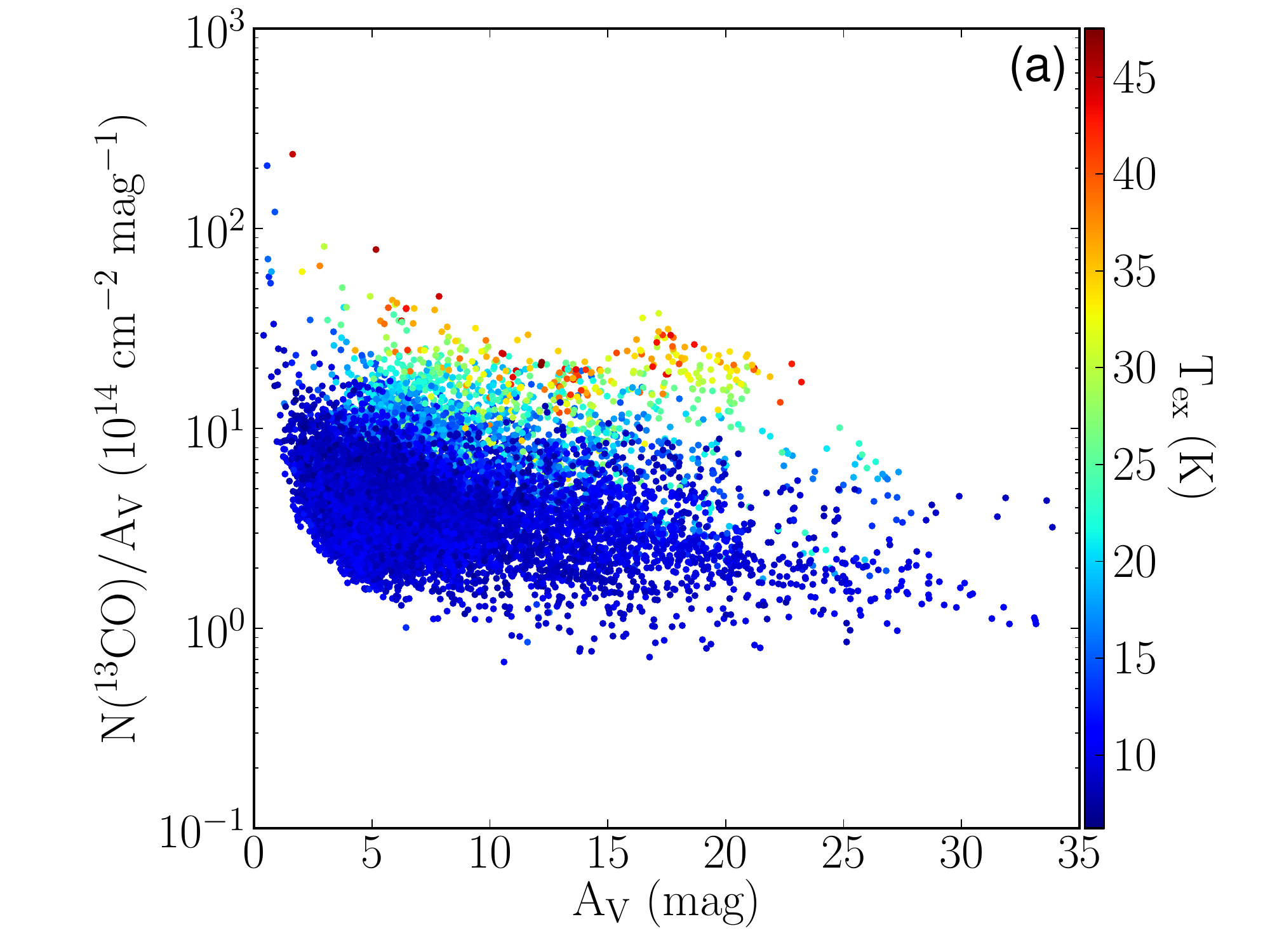}\plotone{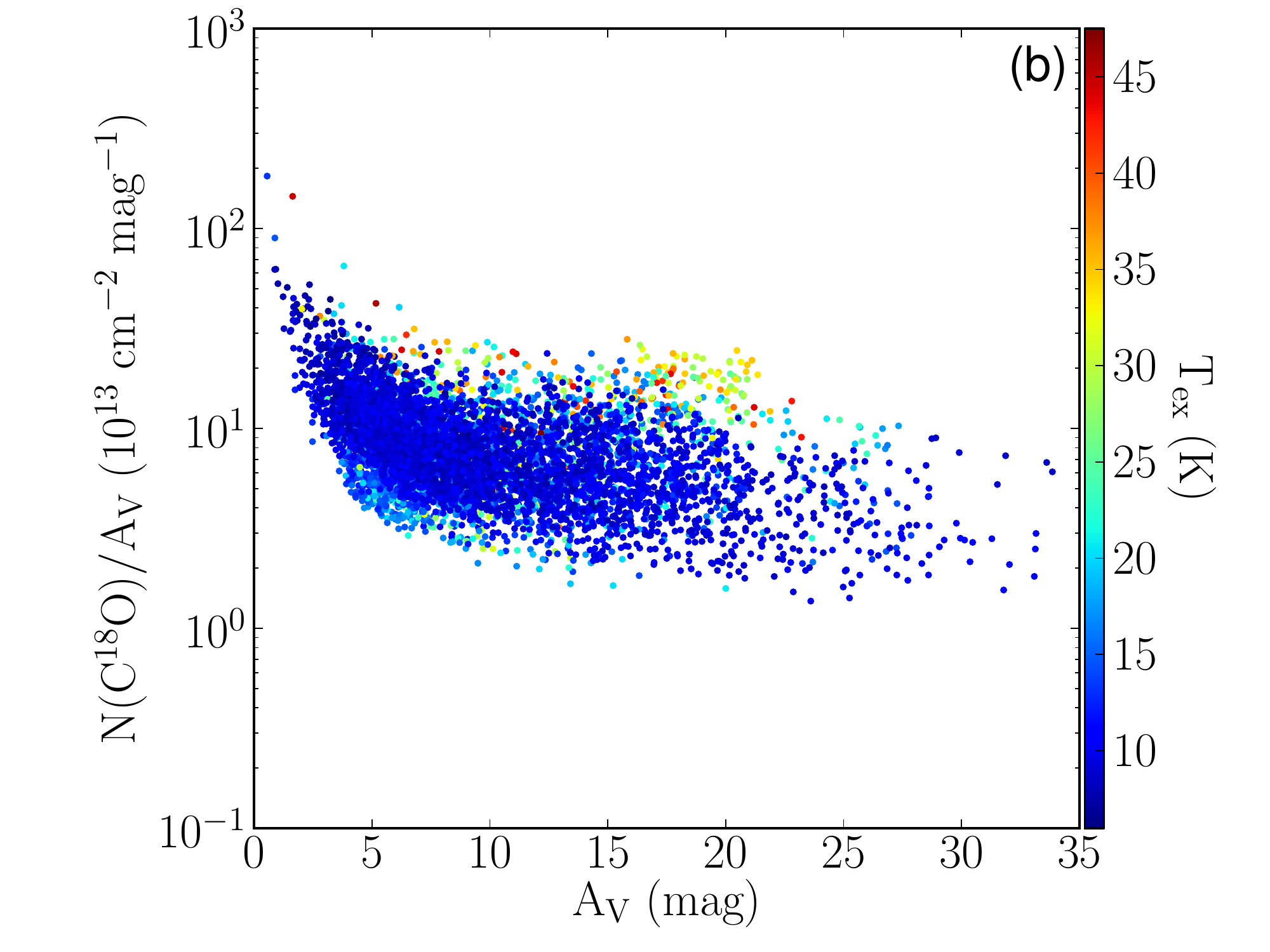}
\caption{
The relation between CO abundance and extinction for (a) \thircons, 
(b) \ceio  over the entire surveyed area. 
Although there is  large scatter, both plots show 
that the abundances of these two rare isotopologues 
clearly decrease with increasing extinction into the cloud. 
This is likely a result of CO depletion in the cold high extinction 
regions of the cloud. The individual points are color-coded 
by the value of the \tex derived from \co 
observations at their position. Note that the hotter pixels have elevated 
abundances compared to the cold pixels and these abundances 
don't appear to decrease with extinction. 
\label{fig:noverav}}
\end{figure*}

In Figure \ref{fig:noverav} we plot the relation between 
(a) \thirco and (b) \ceio abundance and extinction 
in the CMC. Although there is a large scatter in both 
relations, it is clear that the abundances of the two 
species decline with extinction in a manner consistent 
with similar abundance reductions seen in other molecular clouds and
suggestive of CO freeze out onto grain surfaces in the interior 
regions of the cloud. If this decline in abundances at high extinctions
is due to freeze out, we also expect 
the reduction of the gas-phase  abundances to be 
occurring in the coldest regions of the cloud. 
In Figure \ref{fig:noverav} the points are color-coded 
by the gas temperature and we see both a preponderance 
of cold gas in the cloud, and a prominent decrease 
in abundance with extinction for the cold pixels, 
while the warm pixels show no decrease. 
This supports the hypothesis that the decline 
in abundances with extinction in cold regions is largely due to depletion.   
Of course, depletion requires low dust temperature and our analysis 
has been based on measurements of gas temperature. 
Gas temperature generally tracks dust temperature 
\citep[e.g.,][]{2014A&A...568A..27F} but direct observations 
of dust temperature would make for a more compelling case.  
Fortunately, published maps of dust temperature for the CMC 
based on Herschel observations exist \citep{2013ApJ...764..133H}. 
Comparison with our maps shows generally good agreement between 
the gas and dust temperature spatial distributions. In particular, 
in the cold regions away from LkH$\alpha$ 101 
(corresponding to our tiles 01-09 and 15-17, respectively),
the median \td $\sim$ 14.5 K and the minimum
\td $\sim$ 10 K \citep{2013ApJ...764..133H}, far below
pure CO ice sublimation temperature \citep[$\sim$ 17 K,][]{1993prpl.conf..163V}.
Under such conditions, \citet{1995ApJ...441..222B} showed 
a relatively short  CO depletion timescale,  $<$ 10$^6$ yr.
So with \td $\sim$ 10 K and $n_{\rm H_2}$ = 10$^4$ cm$^{-3}$, CO would unavoidably deplete. This is consistent with the reduction of CO abundance we observe in the cold, high extinction regions in the CMC.

At the higher gas temperatures (T $\ga$ 17 K) we would 
expect that thermal evaporation (desorption) of the gas 
from the grain surfaces releases any trapped CO and 
increases its gas phase abundance relative to the cold 
regions of the cloud. Indeed, that is what is clearly shown 
in Figure \ref{fig:abuntex} where the observed abundances
increase with increasing gas temperature. 
The warmest CO gas in our surveyed region is found in 
four contiguous tiles (11-14) that are in close proximity 
to the massive star LkH$\alpha$ 101 and its associated cluster
(shown as the blue circle in Figure \ref{fig:texmap}). 
The CMC dust temperature map from \citet{2013ApJ...764..133H} (see their figure 4)
shows a vast region of high dust temperature (\td $\sim$ 28 K)
near LkH$\alpha$ 101, which coincides with the positions of tiles 11-14 
in our observation. It is in this warm area that the CO abundances 
appear to be at their highest.
CO thermal evaporation is caused by high dust temperature.
For instance, \citet{1995ApJ...441..222B} reported that shortly after a ``star turns on", 
the dust temperature reaches \td = 25 K ($n_{\rm H_2}$ = 10$^4$ cm$^{-3}$, 
T$_{\rm gas}$ = 20 K), and all CO is in the gas phase within 100 yr.  
The abundance enhancement near LkH$\alpha$ 101
is likely caused by hot dust heated by the cluster. Over most of 
 the rest of the cloud  \citet{2013ApJ...764..133H} find 
 \td $\sim$ 10 - 14 K. In these regions the gas temperatures 
 are also found to be near 10 K and a significant amount 
 of CO should be locked in ice mantles on the dust grains 
at least at the highest extinctions \citep[also see][]{1995ApJ...441..222B}. 
 These considerations suggest that the variation of [\thircons] 
 and [\ceions] in the CMC is mostly regulated by the 
 chemistry of depletion and desorption which in turn is driven 
 by the (dust) temperature distribution.
 
Recently, \citet{ripple13} have reported similar results 
for the Orion A molecular cloud. They divided that cloud 
into 9 spatial partitions (1-9) and considered regions 
within those partitions with extinctions, \av $\ga$ 5 magnitudes, 
similar to those studied here for the CMC. 
They found that in regions with 5 $\la$ \av $\la$ 10 
magnitudes, \nthir was generally linearly correlated 
with \avns. However, in those partitions (2, 3 \& 4) that 
also contained higher extinction material and were  
characterized by mean gas excitation temperatures 
that were below $\sim$ 22 K, the \nthirns-\av relation 
became flat above \av $\approx$ 10 magnitudes. 
On the other hand, in two partitions (5 \& 6) with both 
high extinctions and high mean gas excitation temperatures 
(\tex $\sim$ 26-33 K), \nthir continued to rise above 
\av $\approx$ 10 magnitudes. \citet{ripple13} posited 
that this behavior of \nthir with extinction was a result of 
depletion/desorption effects at \av $\ga$ 10 magnitudes. 
Our results for the CMC are certainly consistent 
with those of \citet{ripple13} for Orion A. In this respect, the physical/chemical 
conditions in the Orion and the CMC appear to be quite similar. 
The spatial variations in abundances of the rarer isotopes of CO are largely 
determined by depletion/desorption effects due to spatial variations in dust (and gas)  
temperatures resulting from localized star formation activity within the clouds.

\subsubsection{The {\rm \thircons-to-\ceions} Abundance Ratio: Influence of UV photodissociation}\label{subsubsec:w1318}

In the outer, lower extinction regions of molecular clouds, 
FUV radiation is expected to play a role in cloud chemistry 
\citep[e.g.,][]{1988ApJ...334..771V,2009A&A...503..323V}. 
In low extinction regions (i.e., \av $\la$ 3 mag) 
fractionation of \thirco can increase its abundance relative 
to \cons. At somewhat higher extinctions 
(\av $\la$ 5 mag) selective photodissociation of \ceio is 
likely to occur due to differences in self-shielding of the FUV 
intensity at the dissociation wavelengths for the various CO isotopes.  
The more abundant \co and \thirco isotopes 
are effectively self-shielded at  relatively low extinctions (\av $\la$ 1 magnitudes) 
but  because the dissociation wavelength of \ceio is slightly shifted 
compared to \co and \thircons, and its abundance is relatively 
low so it is not as effectively self-shielded as the main isotopes, 
FUV radiation can penetrate deeper into the cloud and dissociate 
the \ceio \citep{2009A&A...503..323V}. Since our surveys cover relatively high extinction 
regions (\av $\ga$ 5 mag) we would not necessarily expect selective 
photodissociation to strongly effect the \thirco to \ceio abundance ratios.  
However, earlier studies of \citet{1994ApJ...429..694L} and \citet{Shimajiri14} 
suggest that the effects of selective photodissociation may be present 
at much higher extinctions.  

\begin{figure}[htb]
\epsscale{1.2}
\plotone{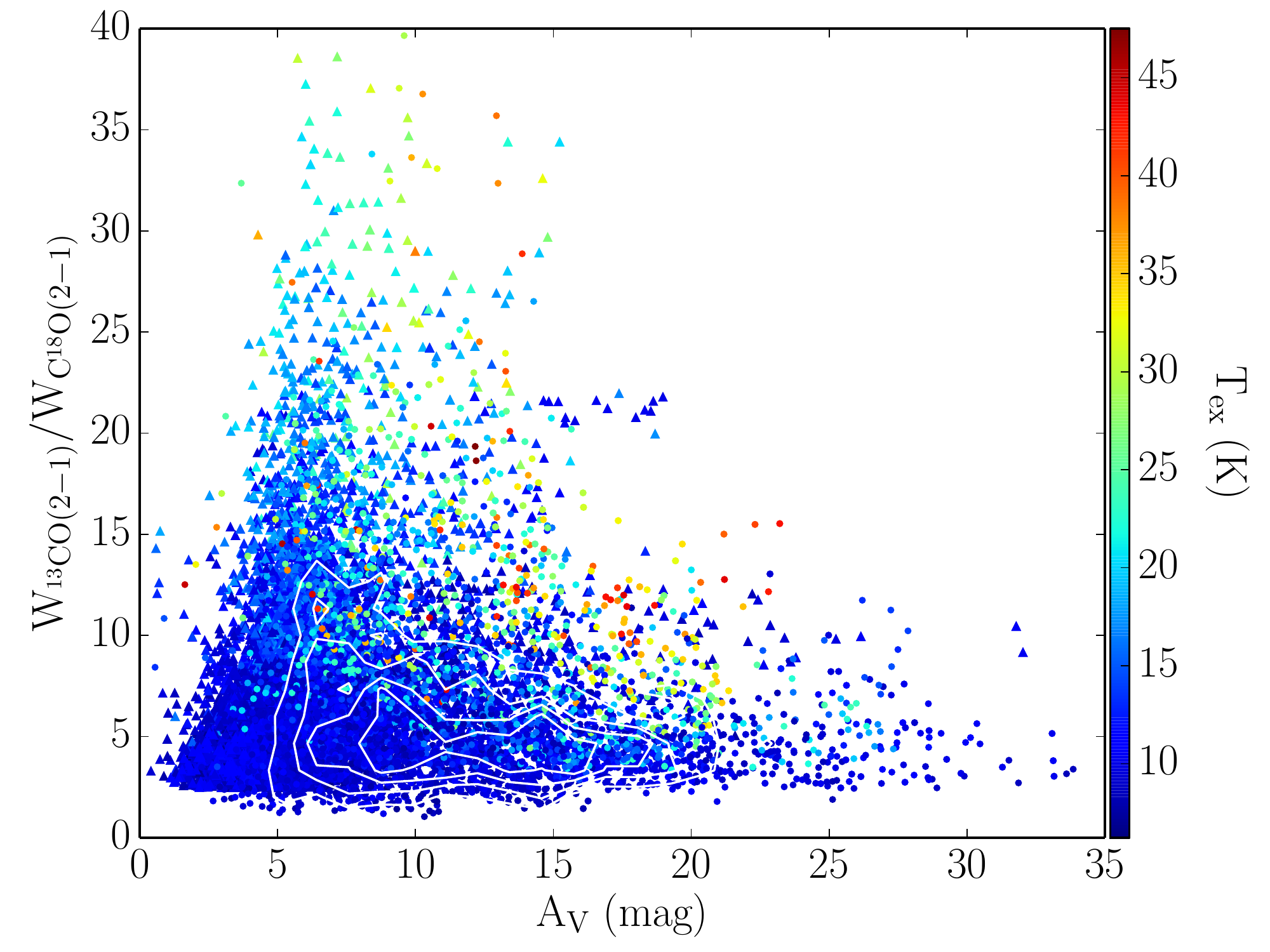}
\caption {
The ratio of W$_{\rm ^{13}CO(2-1)}$ and W$_{\rm C^{18}O(2-1)}$
plotted as a function of \avns. W$_{\rm ^{13}CO(2-1)}$ is everywhere $>$ 5$\sigma_{W,13}$.
Those pixels with W$_{\rm C^{18}O(2-1)}$ $>$ 2$\sigma_{W,18}$ are plotted as
filled circles, those originally with W$_{\rm C^{18}O(2-1)}$ $<$ 2$\sigma_{W,18}$
are set to have  W$_{\rm C^{18}O(2-1)}$ = 2$\sigma_{W,18}$ to reduce scatter in the diagram and are plotted as triangles.
These points are all significant lower limits to the true values of the ratio. 
The white contours show the surface density of pixels with detections.
At low extinctions (\av $\la$ 15 mag) there is 
a large spread in the observed ratios with the largest values 
($\sim$ 30--40) considerably in excess of the solar abundance 
ratio of 5.5. At high extinctions (\av $\ga$ 15 mag) the dispersion in the observed 
ratios is dramatically less and the ratios are close to the solar 
value. See Section \ref{subsubsec:w1318}.
\label{fig:w1318}}
\end{figure}

In Figure \ref{fig:w1318} we plot the ratio of integrated intensities, 
W(\thircons)/W(\ceions), as a function of extinction. This ratio is 
proportional to the abundance ratio for the two species for 
optically thin lines. It has been shown to be sensitive to the 
effects of UV radiation on cloud chemistry and abundances 
\citep[e.g.][]{1994ApJ...429..694L,Shimajiri14}.
For a solar abundance and optically thin lines, the ratio 
should be equal to about 5.5 and one would expect the 
points to scatter around this value, independent of extinction, 
provided the cloud is characterized by a constant relative 
abundance throughout and that the excitation of the two 
species is not too different.  Although at lower extinctions 
(\av $\la$ 15 magnitudes) there indeed appears to be no 
obvious correlation between this ratio and extinction, 
the observed values scatter over an enormous range 
($\approx$ 0 - 40). The highest values of the ratio exceed 
the solar value by nearly an order of magnitude\footnote{We note here 
that in the outer regions of the cloud, \ceio is often 
undetected while \thirco is still quite strong. Because \ceio is in the denominator, 
the ratio is very sensitive to noise in the outer regions. 
However, at all locations \thirco detections exceed the 5 $\sigma$ level. 
To reduce the scatter in the plot due to non-detections of \ceio 
we set the \ceio integrated intensity equal to 2 $\sigma_{W,18}$ 
for those pixels where W(\ceions) is below 2 $\sigma_{W,18}$. We represent the resulting 
lower limits on the calculated ratio as triangles in the plot (see figure).}.  
In contrast, at high extinctions (\av $\ga$ 15 magnitudes) the relation is 
relatively flat with a relatively low dispersion, as might be 
expected for a constant abundance ratio, and is centered 
at a value (approximately 4.5) near but somewhat less than 
the solar value. This behavior does not appear to depend on 
gas temperature (see figure) and thus is not likely a result of 
depletion or desorption processes on grains. It is more likely 
indicative of an active chemical processing of \ceio by FUV 
radiation in regions where \av $\la$ 15 magnitudes. 
The scatter in Figure \ref{fig:w1318} is so large at these 
extinctions because the relative abundance of \thirco 
and \ceio is highly unstable possibly due to the stochastic 
variations in FUV flux resulting, in turn, from such factors as the 
cloud structure and geometry and nature of the external radiation field. 

\begin{figure}[htb]
\epsscale{1.1}
\plotone{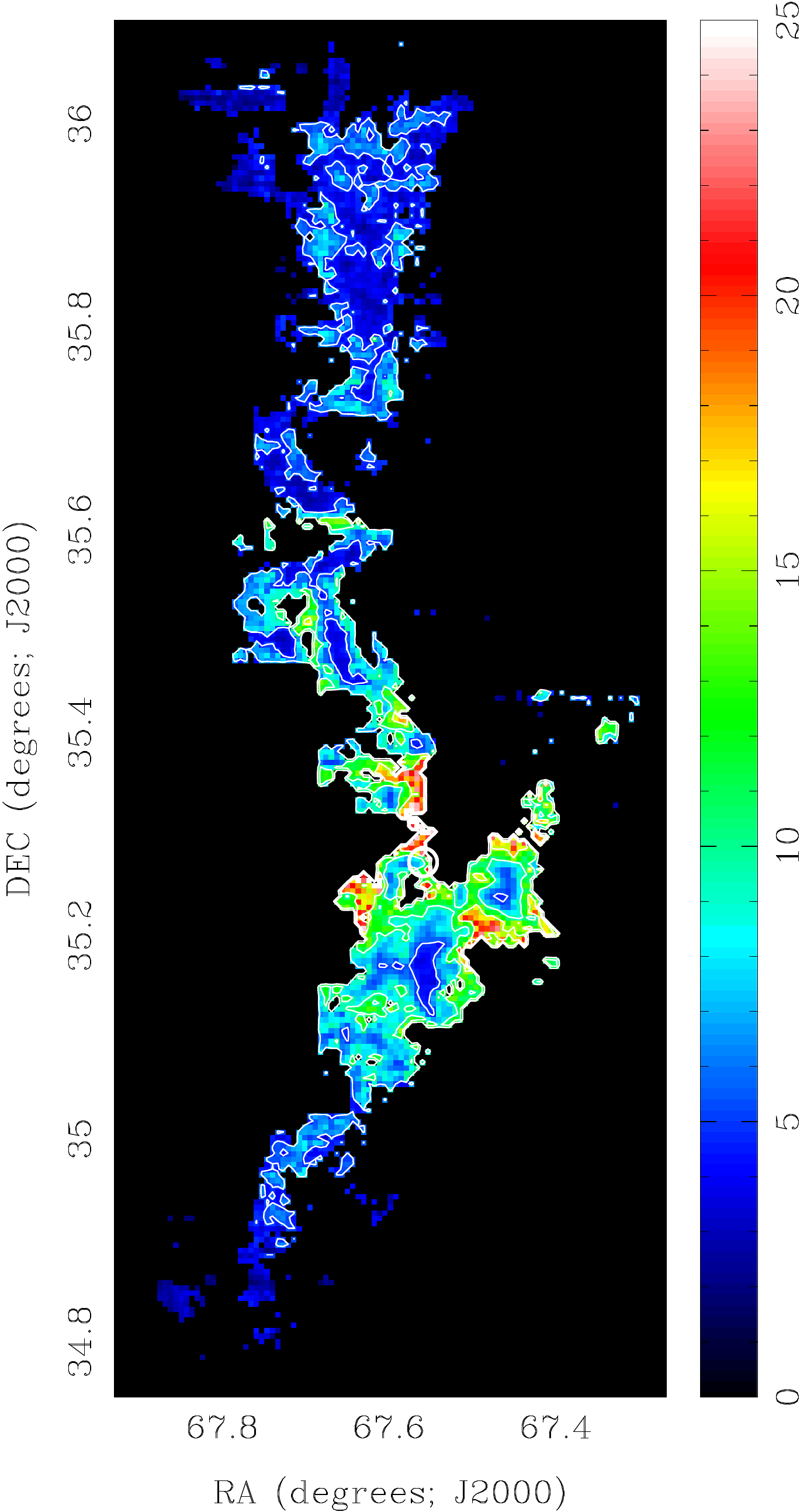}
\caption {
A map of the  \thirco to \ceio integrated intensity 
ratio in the southern portion of the surveyed region. 
The contours show W(\thircons)/W(\ceions) = 5, 10, 15.
The largest values of this ratio appear in the outer regions 
where the chemistry is driven by FUV radiation that 
penetrates deeply into the cloud. It is clear that this 
effect is larger in the southern region of the map 
(i.e., $\delta \approx 35\arcdeg10\arcmin$ to $35\arcdeg20\arcmin$) 
which is in close proximity to the embedded cluster (open white circle)
and its UV radiation field than in the more remote northern 
portion of the region.
\label{fig:w1318map}}
\end{figure}

In Figure \ref{fig:w1318map} we show a map of the distribution 
of the integrated intensity ratio over the southern portion of 
the CMC. Examination of this map shows systematically 
increasing values of the integrated intensity ratio at the edges 
of the cloud with the material nearest the star LkH$\alpha$ 101 
exhibiting the largest center to edge increases in the ratio. 
This latter region appears to contain a well developed photon dominated region (PDR)
while the dense cloud material to the north is much more 
quiescent in this regard. In their study of the Orion cloud  
\cite{Shimajiri14} reported averaged values of the \thirco 
to \ceio integrated intensity ratio ranging from about 10 to 17 
in the seven PDRs they observed 
in that cloud. Moreover they also found the variations in this 
ratio to be linked to variations in the external FUV radiation field. 

In the inner, higher extinction regions of the CMC, where FUV 
photons cannot penetrate, the abundances are more 
chemically stable and apparently somewhat below the solar 
value. Our observations do indicate, however, that FUV 
photons are present at relatively large (projected) depths 
(\av $\sim$ 15 magnitudes) in the cloud confirming results 
from earlier studies \citep[e.g.][]{1994ApJ...429..694L,Shimajiri14}.

We now address the earlier observation that the relations between the 
CO abundances and temperature (Figure \ref{fig:abuntex}) show positive 
correlations but over different temperature ranges. One might have 
expected both \thirco and \ceio molecules to show abundances that increased with 
temperature over similar temperature ranges, since their binding 
energies are very similar. However, the difference may result from 
the effects of selective UV dissociation of \ceions. 

\begin{figure}[htb!]
\epsscale{1.2}
\plotone{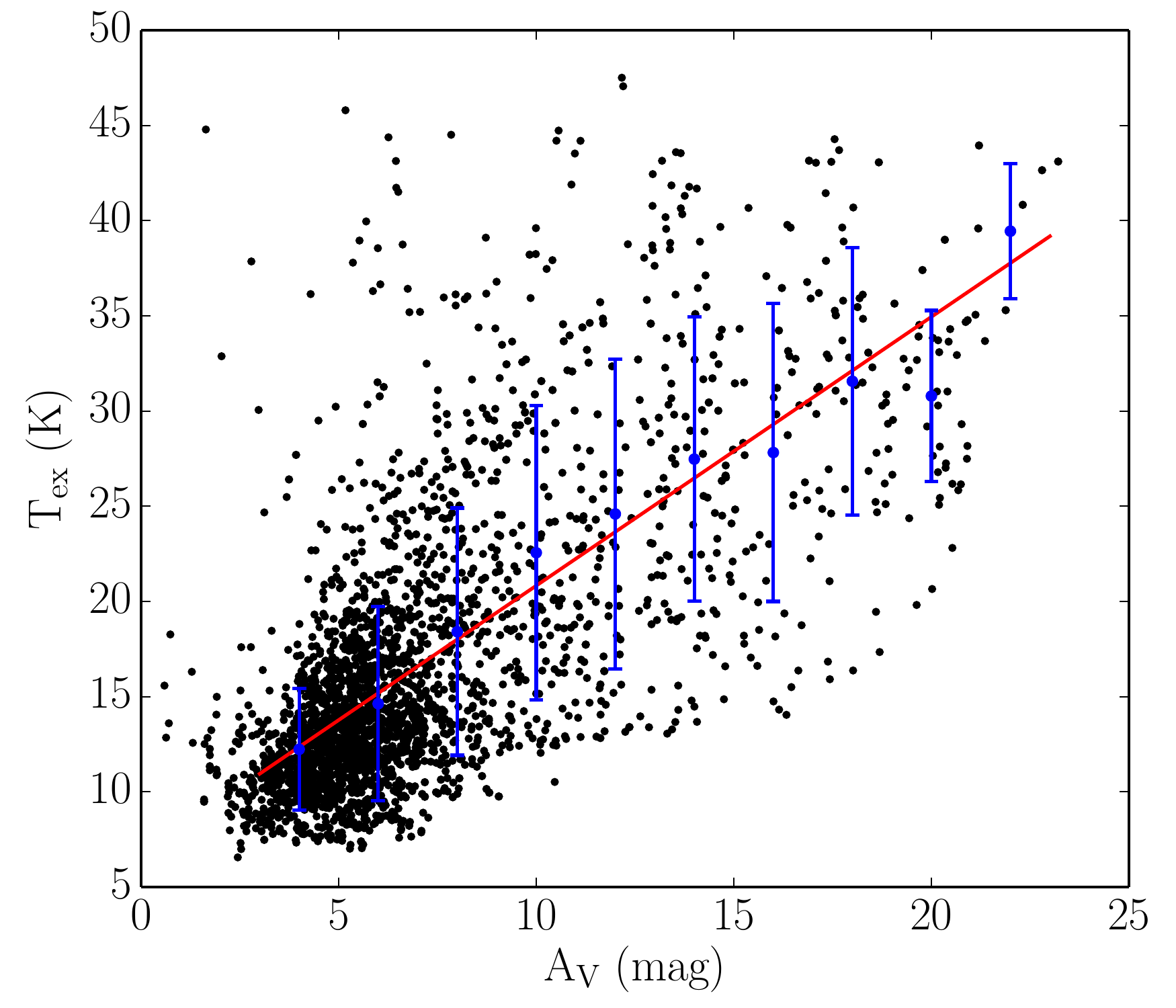}
\caption{
\texns-\av relation for hot tiles 11-13. 
The width of the bins is 2 mag. The red solid line shows
linear fitting on the unbinned data points.
This shows the correlation between
\tex and \av in the region near LkH$\alpha$ 101.
Typical errors in \tex are always below 5\%. Typical uncertainties in \av are between 0.05 and 0.1 magnitudes. See text.
\label{fig:texavhot}}
\end{figure}

In the CMC we also find \tex to be correlated with \av in regions 
where warm gas is present. Figure \ref{fig:texavhot} displays the relation between
\tex and \av for the warm tiles (11-13) in our map. 
Although the scatter is large at high \avns, there is no doubt of a 
strong trend between the two quantities with \tex $\propto$ \avns. 
Interestingly in this region the temperature reaches the CO sublimation 
temperature at roughly \av $=$ 6-7 magnitudes. Therefore it is possible that, 
even though \ceio is being evaporated from grains at \av $\ga$ 6 magnitudes, 
(where \tex $\ga$ 17 K) it is also being selectively photodissociated by deeply penetrating 
UV radiation until depths of \av $\sim$ 15 magnitudes. 
For larger values of \avns, dust absorption effectively removes the dissociating 
UV radiation and higher gas (dust) temperatures associated with this region liberate 
more \ceio from grain surfaces. Indeed, the \ceio abundances 
at these higher extinctions and temperatures may reach levels that also 
allow some \ceio self-shielding. Consequently, \ceio does not reach the 
expected levels of increased abundance due to evaporation from grains 
until cloud depths are sufficient to shield the molecule from UV radiation. 
Because of its higher abundance, \thirco self-shields at much lower 
cloud depths and the effects of its increased abundance, due to evaporation 
from grains, are observed at lower extinctions (and corresponding lower temperatures) than \ceions.

\subsection{X(CO) on Sub-Parsec Scales}\label{subsec:wav}

Even though H$_2$ is the dominant constituent of molecular 
clouds, \co emission is the most accessible tracer of 
such gas in the ISM. It is often the only molecular tracer that 
can be readily detected in distant clouds and galaxies and 
consequently has been used to estimate the mass of the 
molecular component of the ISM in these systems 
\citep[e.g.][]{bolatto2013}. To make such estimations requires 
knowledge of the so-called X-factor, that is, the CO conversion factor, 
\begin{equation}\label{eq:xco}
\rm X_{CO} 
= N(H_2)/W(CO_{J=1-0})
\end{equation} 
The standard determination of this factor is via use of  virial theorem techniques
\citep[e.g.][]{1987ApJ...319..730S}. However, the most straightforward 
way to determine the X-factor is by 
comparing direct measurements of CO and \av on sub-parsec scales in local clouds. 
Traditionally this is accomplished by measuring the slope of the correlation 
between the two quantities at low extinctions (\av $<$ 5 mag) 
where the CO appears to be effectively thin 
\citep[e.g.][]{dickman78,frerking82,2006A&A...454..781L}. 
We adopt a slightly different approach and use all our observations to compute 
a global or average X-factor for the CMC cloud, that is, $<$$\rm X_{CO(1-0)}$$>$ 
= $<$$\rm N(H_2)$$>$/$<$$\rm W(CO)$$>$ $=$ 2.53$\times$10$^{20}$ cm$^{-2}$  
(K km s$^{-1}$)$^{-1}$. This value is 25\% larger than the standard Milky Way 
value but within the range of values derived for individual molecular clouds in 
other studies \citep[e.g.][]{bolatto2013}. Another approach is to derive the on-the-spot  \xcons, that is, we evaluate Equation \ref{eq:xco} at each 
pixel in our map and obtain 
$<$$\rm X_{CO(1-0)}$$>_{\rm ots}$ 
= $<$$\rm N(H_2)/W(CO)$$>$ $=$ 3.1$\times$10$^{20}$ cm$^{-2}$  
(K km s$^{-1}$)$^{-1}$. The former $<$\xcons$>$ is equivalent to the latter weighted by the integrated intensity.

However, both recent observational studies \citep{pineda08,pineda10,bieging10,ripple13,2014ApJ...784...80L} 
and numerical simulations \citep{2011MNRAS.415.3253S,2011MNRAS.412.1686S}
have shown that \xco can be strongly dependent on local physical conditions. 
Since the CMC displays a 
wide range of physical conditions (e.g. temperature, star formation activity and spatially variable 
CO abundances), it would be interesting to measure both  
\xco and its internal variation on sub-cloud, sub-parsec scales. 

\begin{figure}[htb!]
\epsscale{1.2}
\plotone{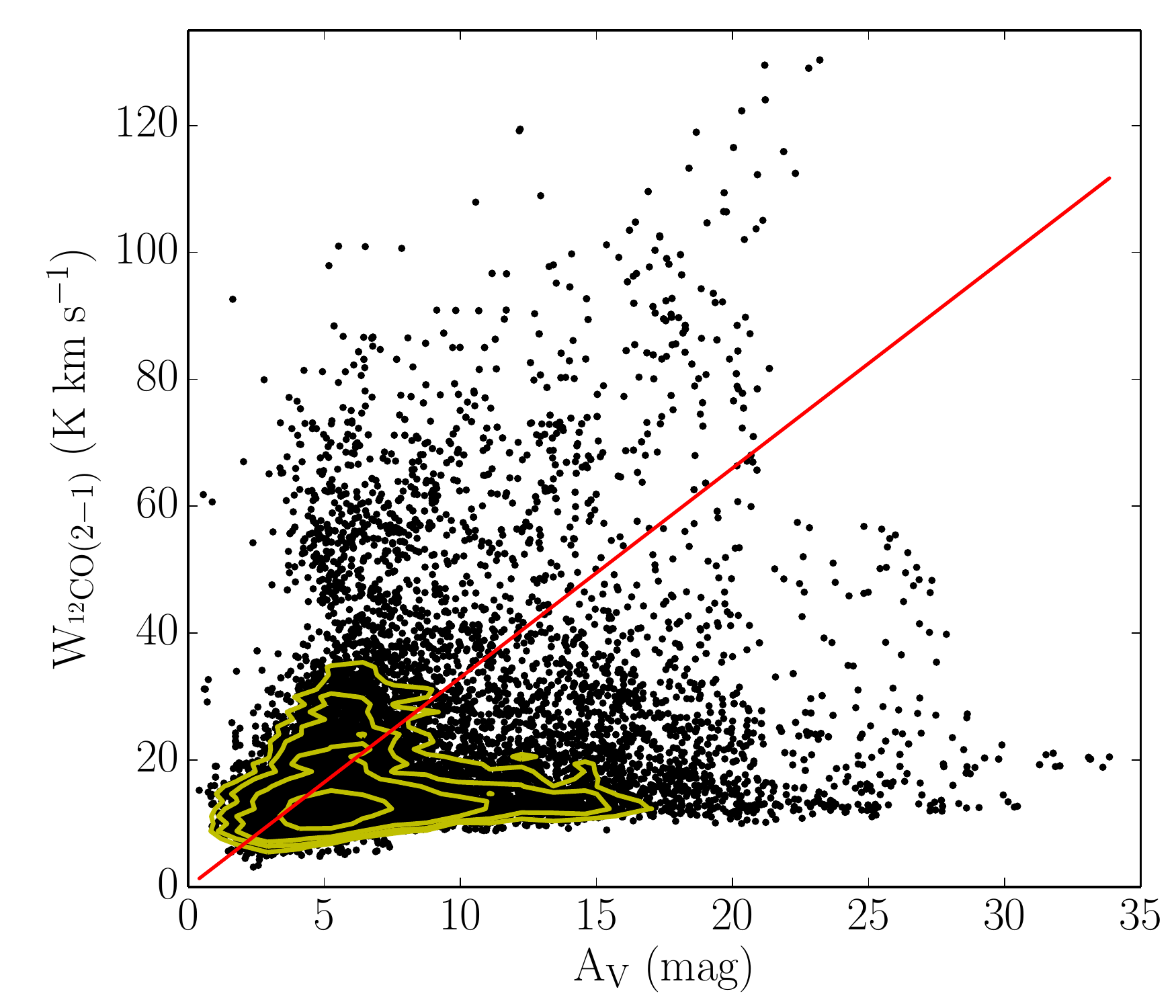}
\caption{A pixel-by-pixel comparison between \cons(2-1) 
integrated intensity and visual extinction \av for the 
entire surveyed area. Each point represents a 19$\arcsec$ pixel which 
corresponds to $\sim$ 0.04 pc at a distance of 450 pc. The red 
solid line represents the standard Milky Way \xco with 
a slope of 3.3 (K km s$^{-1}$) mag$^{-1}$. The yellow contours
show the surface density of points. Uncertainties are comparable to 
or less than the size of the symbols. See Section \ref{subsec:wav}.
\label{fig:wav12}}
\end{figure}

To do this we first performed the traditional pixel-by-pixel comparison between 
W$_{\rm ^{12}CO(2-1)}$ and \av for the entire 
surveyed region, as shown in Figure \ref{fig:wav12}. 
Assuming a constant gas-to-dust ratio  
N(H$_2$)/\av=9.4$\times$10$^{20}$ cm$^{-2}$ mag$^{-1}$   
\citep{bohlin78,rachford2009,pineda08} and a constant 
line integrated intensity ratio 
W$_{\rm CO(2-1)}$/W$_{\rm CO(1-0)}$ = 0.7, X$_{\rm CO(1-0)}$ is inversely
proportional to W$_{\rm ^{12}CO(2-1)}$/\av in the plot. 
In the Milky Way disk, the typical value is 
X$_{\rm CO(1-0)}$ $\sim$ 2$\times$10$^{20}$ cm$^{-2}$ 
(K km s$^{-1}$)$^{-1}$ \citep[e.g.,][]{bolatto2013}. 
This implies a reference value \avns/W$_{\rm CO(2-1)}$ 
$\approx$ 0.30 (K km s$^{-1}$)$^{-1}$ mag. We plot a red line 
indicating this reference value in Figure \ref{fig:wav12}, with a slope 
of W$_{\rm CO(2-1)}$/\av $\approx$ 3.3 (K km s$^{-1}$) 
mag$^{-1}$. At fixed \avns, a point above this line indicates 
a smaller X$_{\rm CO(2-1)}$, and vice versa. 

\begin{figure*}[htb!]
\epsscale{1.1}
\plotone{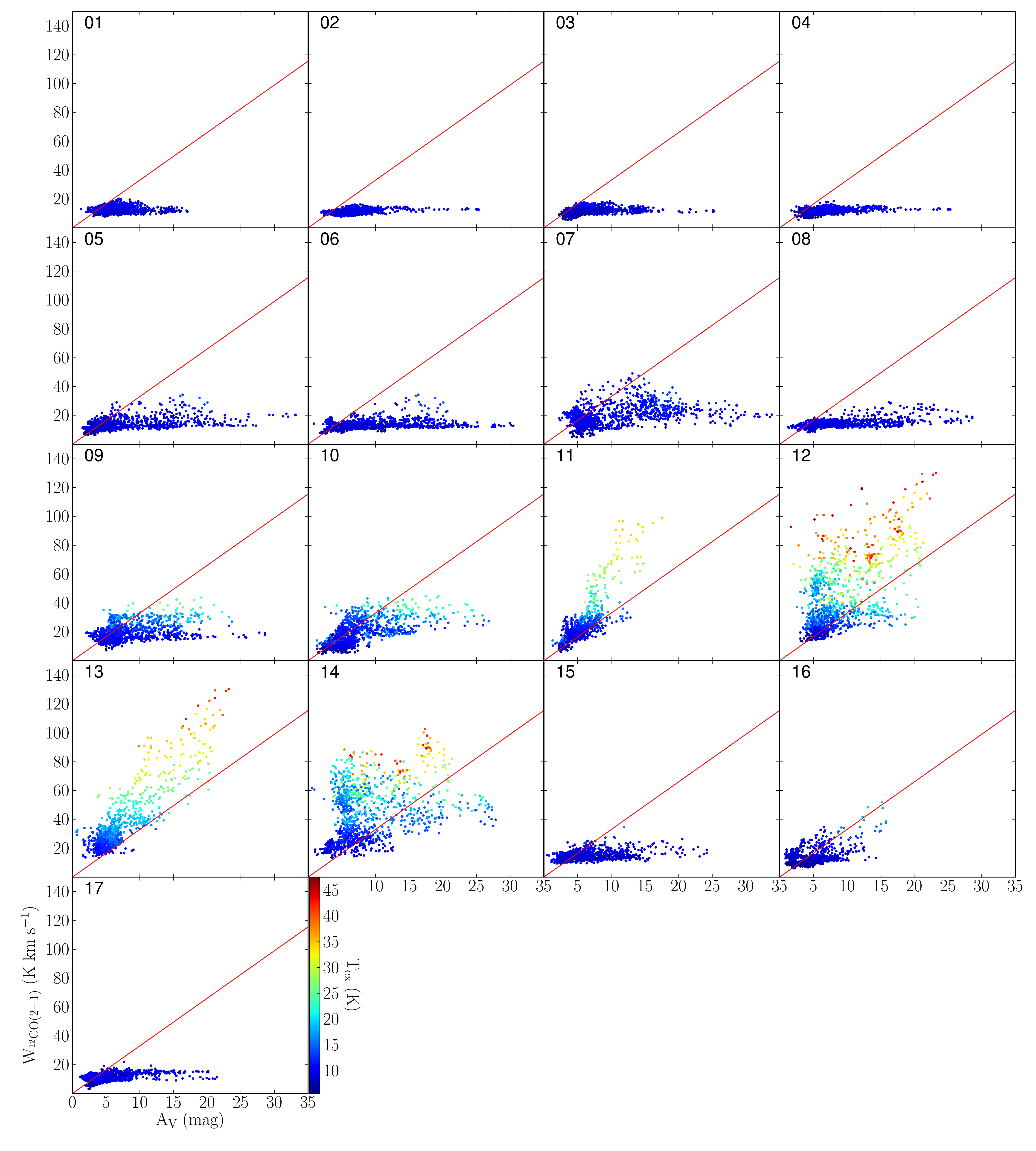}
\caption{A pixel-by-pixel comparison between \co 
integrated intensity and \av within each tile. All tiles 
are shown in the same scale in both axes.  
The red solid lines represent the standard Milky Way 
\xco with a slope of 3.3 (K km s$^{-1}$) mag$^{-1}$.}
\label{fig:twav12}
\end{figure*}

We can see significant scatter in this plot indicating significant spatial 
variations in X$_{\rm CO(2-1)}$ on pixel or sub-parsec spatial scales. 
In order to investigate the spatial dependence of X$_{\rm CO(2-1)}$, 
we determined the W$_{\rm ^{12}CO(2-1)}$--\av 
relation in each individual tile of our CO survey, following 
Section  \ref{subsec:tilewav}. These relations are shown in 
Figure \ref{fig:twav12}. There are several familiar features of 
these plots that are similar to what we 
have seen in Figure \ref{fig:ncol13}, i.e., less scatter in the relations within the individual tiles, 
systematic spatial variations in  the W$_{\rm ^{12}CO(2-1)}$ -- \av  
relation across the cloud, and good correlation  
between high W$_{\rm ^{12}CO(2-1)}$ and high gas temperature 
(\tex $\ga$ 15 K). 

The general similarity\footnote{
By ``similarity" we mean the hot tiles 11-14 show steeper slopes  than
the cold tiles in both W$_{\rm ^{12}CO(2-1)}$ - \av (Figure \ref{fig:twav12})
and N(\thircons) - \av (Figure \ref{fig:ncol13}) relations. The
apparent relative differences between the the cold tiles in 
Figure \ref{fig:ncol13} compared to Figure \ref{fig:twav12} is a 
result of different y-axis scaling.}~
in the behavior of W$_{\rm ^{12}CO(2-1)}$ 
and N(\thircons) with extinction is perhaps surprising given that 
the \co emission is very optically thick while \thirco 
emission is considerably less thick and mostly optically thin. 
Since \nthir is proportional to W$_{\rm ^{13}CO(2-1)}$,
consider that in this situation $\rm W_{^{13}CO(2-1)}$ 
$\approx [J_\nu(T_{ex}) - J_\nu(T_{bg})] \times  \tau^{13}_\nu \Delta v_{13}$, 
while  $\rm W_{^{12}CO(2-1)}$ $\approx  [J_\nu(T_{ex}) - J_\nu(T_{bg})] \times \Delta v_{12}$, 
where $\Delta v$ is the line width of the corresponding spectrum. 
Unlike W$_{\rm ^{13}CO(2-1)}$, W$_{\rm ^{12}CO(2-1)}$ has no dependence 
on $\tau$ or column density, but it is proportional to \texns, 
the gas excitation temperature. 

\begin{figure}[htb!]
\epsscale{1.2}
\plotone{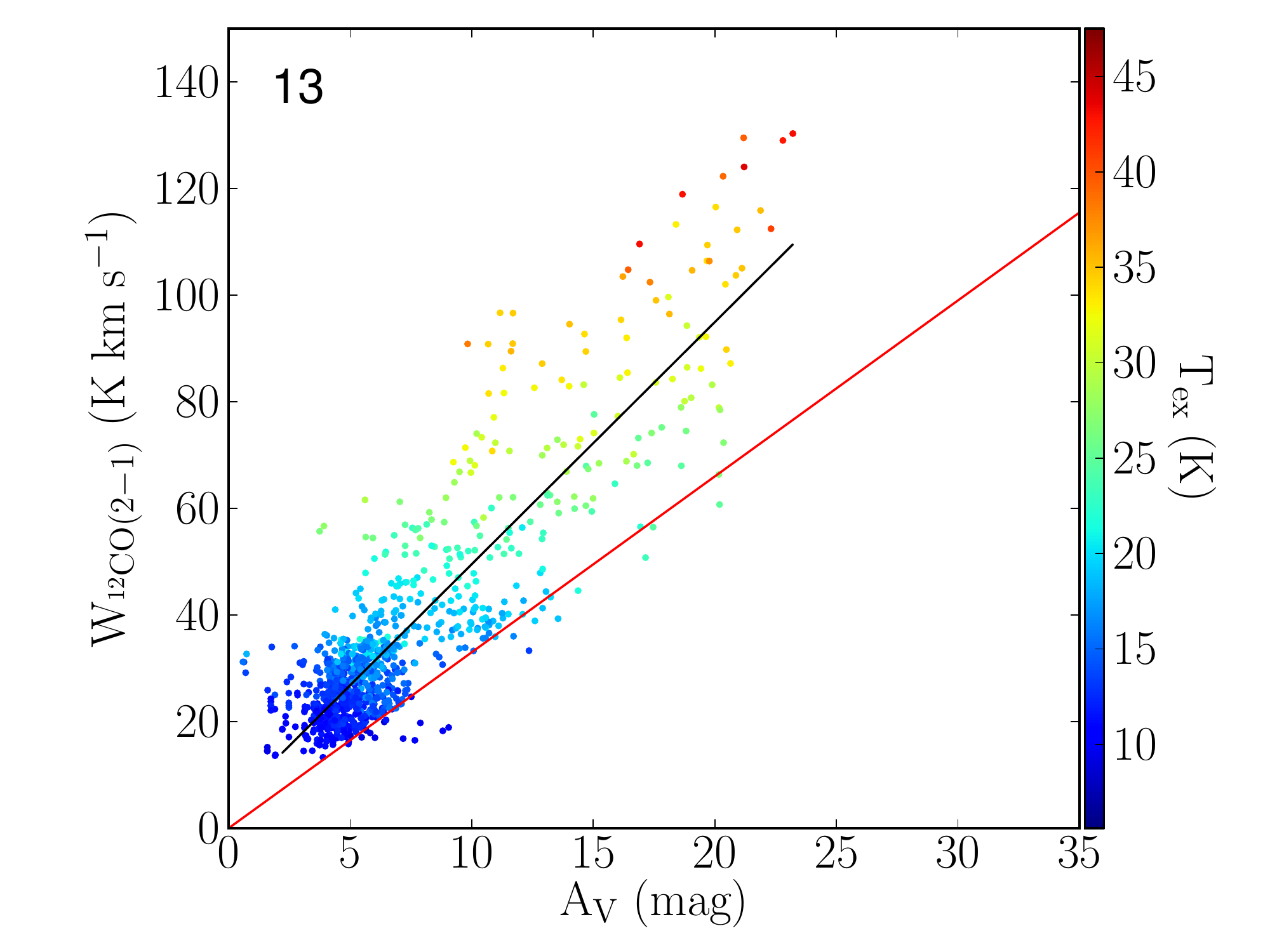}
\caption{Tile 13 from Figure \ref{fig:twav12}. The black solid line is a 
linear regression. The fitted slope corresponds to \xco = $1.5\times10^{20}~{\rm cm}^{-2}~({\rm K~km~s}^{-1})^{-1}$. 
The red solid line represents the fiducial Milky Way value, \xco = $2.0\times10^{20}~{\rm cm}^{-2}~({\rm K~km~s}^{-1})^{-1}$.  
\label{fig:wav12tile13}}
\end{figure}

Particularly interesting in this regard is the behavior of 
W$_{\rm ^{12}CO(2-1)}$ in tile 13 near where the embedded 
cluster is heating the cloud. Here, there appears to be a 
clear linear relation between W$_{\rm ^{12}CO(2-1)}$ and \av to the highest 
measured extinctions ($\approx$ 25 magnitudes) despite the 
fact that the \co emission in this region of the cloud is 
so saturated that it is self-reversed (see Figure \ref{fig:spec}).  
Because the CO emission is so optically thick, this behavior 
must be primarily the result of the gradient of \tex with \av  observed in this 
region of the cloud (see Figure \ref{fig:texavhot}) . 
This is also evident in
Figure \ref{fig:wav12tile13} which shows an expanded view of the W$_{\rm ^{12}CO(2-1)}$--\av relation for tile 13.  
We can clearly see a temperature gradient in the distribution of color-coded points from low  to high \avns. 
Both W(CO) and T increase by a factor of $\sim$ 5-6 from low to high \avns. 
This indicates that the temperature gradient accounts for the bulk of the change in W(CO) with \av in this region. 
Of course, other factors could contribute to the increase in W(CO) such as 
an increase in velocity dispersion with temperature, or even increased [CO] 
due to desorption as with \ceio and \thirco as discussed previously. 
We do find some evidence for a marginal (35\%) increase in the velocity 
dispersion of the hot gas, but its magnitude is far from sufficient to account 
for the increased ($\times 6$) measures of W(CO). A desorption induced 
increase in CO abundance most certainly has occurred in the hot gas, 
and although the \co line is very optically thick, we might expect that additional 
photons could leak out in the line wings  and possibly increase the measured 
velocity dispersion. But as we already remarked,  this effect cannot be significant 
since the change in peak brightness is observed to be comparable to the change 
in W(CO) and substantially larger than changes in the velocity dispersion.

Consequently, in the set of tiles near the massive star LkH$\alpha$ 101 
(11-14) we conclude that  heating of CO enhances W$_{\rm ^{12}CO(2-1)}$ 
in a manner that induces a positive correlation with extinction despite the fact that the emission is saturated. 
These observations suggest that heating by OB stars produces gradients in
the excitation temperature of the CO emitting gas such that in warm regions 
W(\cons)  is proportional to \av and thus to N(H$_2$) over a wide range of extinctions. 
This, in turn, indicates that a single, meaningful X-factor can be derived for 
these hotter regions using the observations. In Figure \ref{fig:twav12}, tile 13 
shows the best correlation between W(\cons) and \avns. 
We performed a linear regression for tile 13, which is shown in 
Figure \ref{fig:wav12tile13}. Our result implies 
$$X_{\rm CO} \simeq 1.5\times10^{20}~{\rm cm}^{-2}~({\rm K~km~s}^{-1})^{-1}$$
in this tile over the range 3 mag $<$ \av $<$ 25 mag. 
From the average of the individual, on-the-spot, \xco values in tile 13 
we find $<$\xcons$>$ $=$ $\rm 1.3 \times 10^{20}~cm^{-2}~(K~km~s^{-1})^{-1}$, 
in excellent agreement with the value derived from the linear regression. 
This clearly supports the idea of a constant value in the hot gas.
The derived value is less than both that found for the global average of the CMC and 
the value ($\rm 2.0\times10^{20}~cm^{-2}~(K~km~s^{-1})^{-1}$) 
usually adopted for the Milky Way.

This result would suggest that in regions heated by OB stars 
(e.g., as HII regions, extragalactic starbursts, etc.), reasonably 
accurate gas masses could be derived using a standard X-factor 
analysis with a single value of the X-factor. Unfortunately it is 
not at all clear how one would predict the appropriate value to use 
in all situations. 

\begin{figure*}[htb!]
\epsscale{0.55}
\plotone{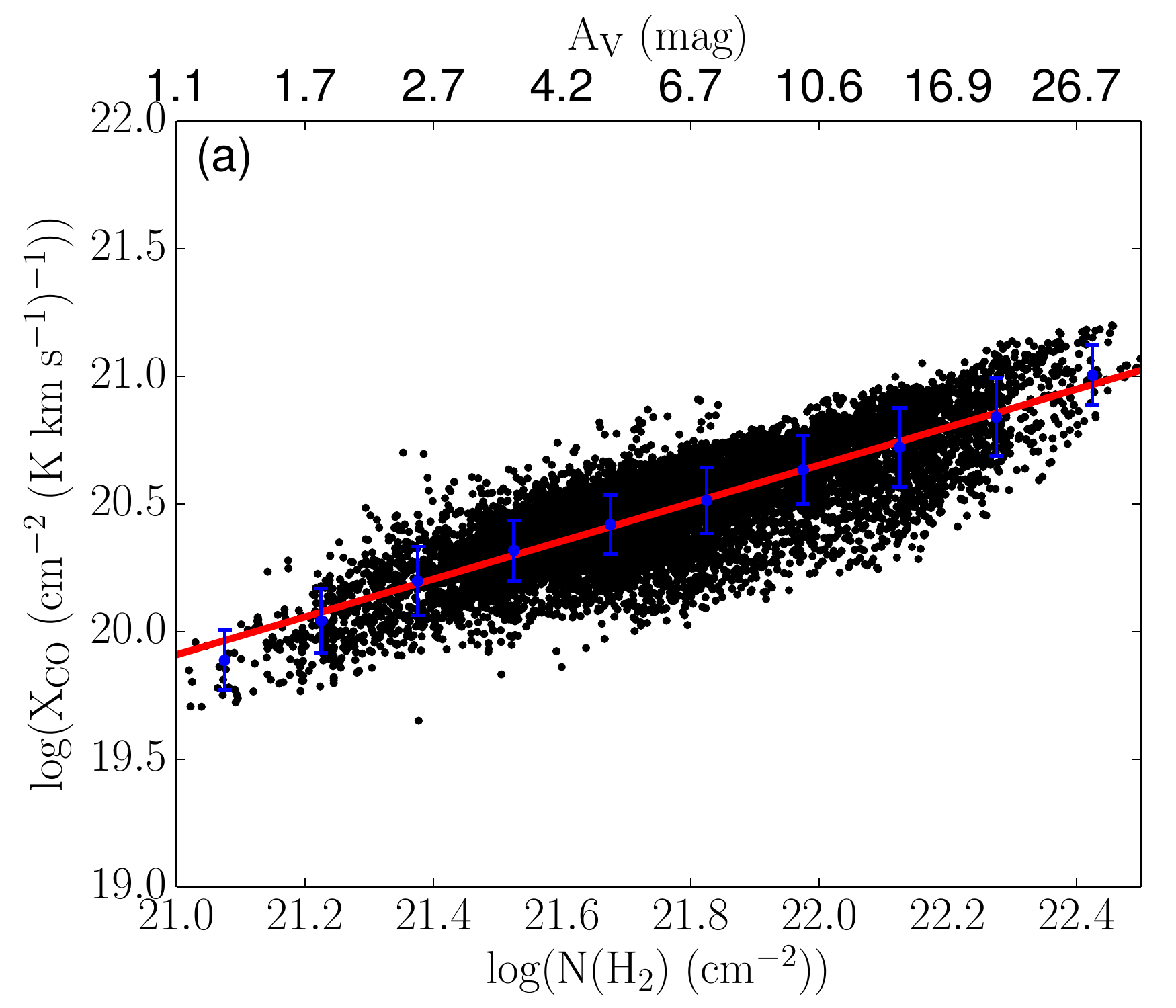}\plotone{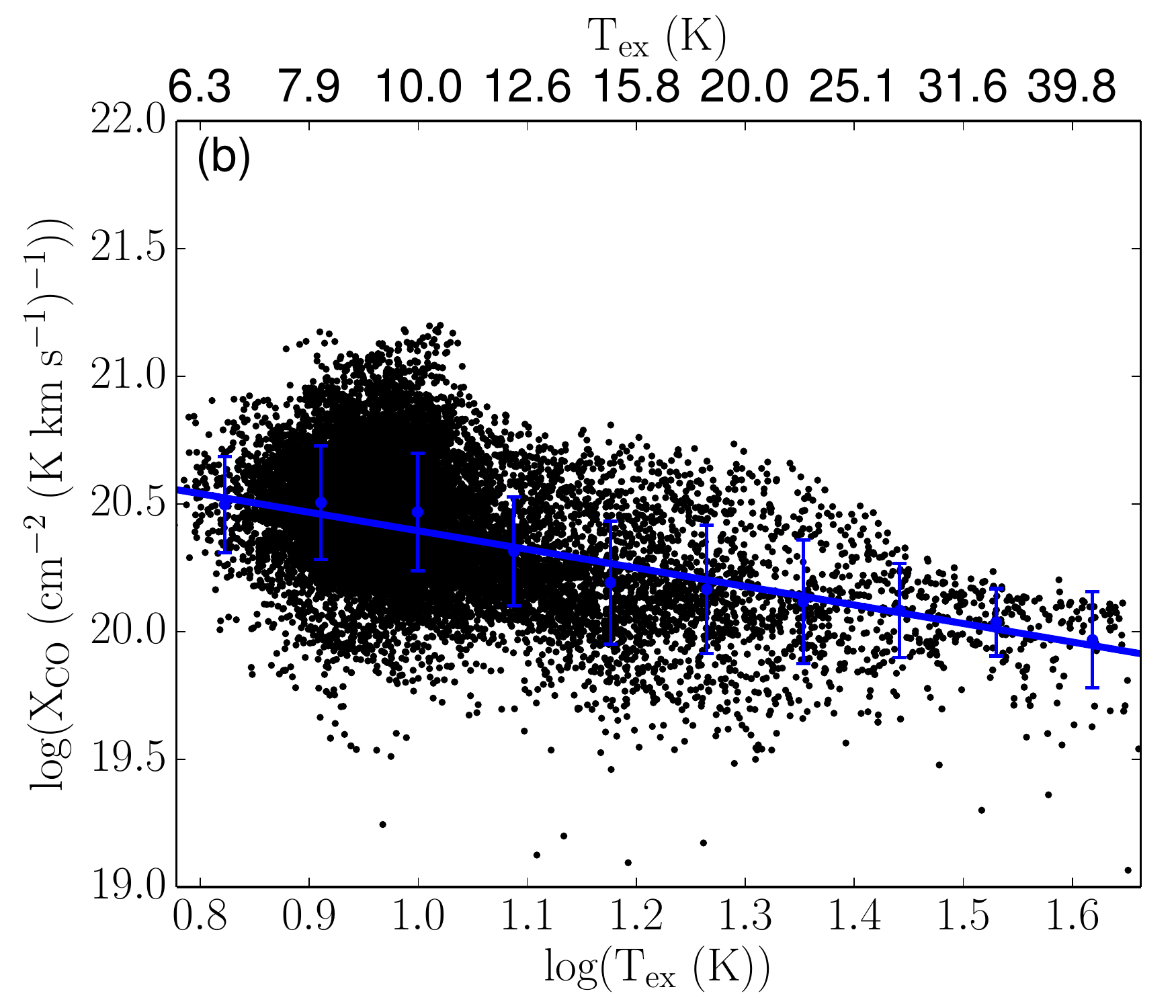}
\caption{
Relations between (a) \xco and N(H$_2$) in cold tiles (excluding tiles 11-14), 
and (b) \xco and \tex in the entire California Cloud. 
Red (a) and blue (b) lines represent least-square fits to unbinned and binned (every 0.09 dex) data, respectively. The typical uncertainties in the X factor are between 5-7\%.
\label{fig:XcovsAv}}
\end{figure*}

However some clues may be found in the analysis of the regions  
away from OB stars, where there is only ambient heating of the 
gas from cosmic rays and the general galactic UV radiation field. 
Here we observe no correlation between W$_{\rm ^{12}CO(2-1)}$ 
and \av at modest to high extinctions (i.e., \av $\ga$ 3-5 mag). 
This behavior is  due to the fact the \co line is almost always 
saturated at high extinctions and the cold gas is very nearly isothermal. 
Consequently, these  cold, quiescent regions, are not characterized by 
a single empirical X-factor, instead, the X-factor systematically increases 
with extinction and, in the CMC, is almost always larger than the Milky Way 
value. This is shown in Figure \ref{fig:XcovsAv}a where we plot the relation 
between the on-the-spot X-factor and \av for the regions of the cloud not 
heated by the cluster (i.e., tiles 1-10 and 15-17). As expected there is a clear 
correlation and a linear least-squares fit to the unbinned data finds that 
\xco $\propto$ N(H$_2$)$^{0.7}$. For the cold tiles we find 
$<$\xcons$>$ =  3.4 $\times$ 10$^{20}$~${\rm cm}^{-2}~({\rm K~km~s}^{-1})^{-1}$. 
The (constant) X-factor in the hotter regions (e.g., tile 13) has a value considerably 
lower than those in the cold regions, suggesting that \xco may also increase with 
decreasing gas temperature within the cloud. Such a trend is clearly 
evident in our observations and is shown in Figure \ref{fig:XcovsAv}b where 
we plot the relation between the on-the-spot \xco and \tex for the 
entire cloud.  A linear least-square fit to the (binned) data gives, 
\xco $=$ 2.0 $\times$ 10$^{20}$(\texns/10)$^{-0.7}$~${\rm cm}^{-2}~({\rm K~km~s}^{-1})^{-1}$. 
In this instance we chose to fit the binned data to give greater weight to high temperature measurements.
A fit to the unbinned data gives a slightly steeper slope (-0.9 vs. -0.7).

The observed inverse correlation between \xco and \tex is potentially interesting. 
It suggests that differences between globally averaged X-factors derived for 
different clouds could be a result of variations in the relative amounts of hot 
and cold gas within the clouds, provided their velocity dispersions are similar. 
Active star forming clouds with OB stars and HII regions, such as Orion, 
might be expected to have average X-factors that are lower than more 
quiescent clouds, such as the CMC. Indeed, 
\cite{1999ApJ...520..196D} derive \xco $=$ 1.35 $\times$ 10$^{20}$
${\rm cm}^{-2}~({\rm K~km~s}^{-1})^{-1}$ for the entire Orion complex from 
$\gamma$ ray data, in agreement with recent determinations using CO \citep{ripple13}. 
Moreover, $\gamma$ ray observations of Orion with the Fermi satellite by 
\citet{2012ApJ...756....4A} found that in the (cold) regions removed from the 
OB stars and HII regions, \xco $=$ 2.2 $\times$ 10$^{20}$
${\rm cm}^{-2}~({\rm K~km~s}^{-1})^{-1}$, while in the (hotter) parts of 
the cloud complex associated with the HII regions, \xco $=$ 1.3 $\times$ 
10$^{20}$~${\rm cm}^{-2}~({\rm K~km~s}^{-1})^{-1}$, very similar to what we find in the California cloud. 
For the Taurus cloud, which has no OB stars or HII regions, \citet{pineda10} 
find \xco = 2.1 $\times$ 10$^{20}$~${\rm cm}^{-2}~({\rm K~km~s}^{-1})^{-1}$ 
from CO observations of that cloud. A relatively high value (2.54 $\times$ 
10$^{20}$~${\rm cm}^{-2}~({\rm K~km~s}^{-1})^{-1}$ ) for \xco  was also 
derived from Planck observations of high latitude diffuse and presumably cold, 
CO gas \citep{2011A&A...536A..19P}.  
We noted earlier that the global X-factor for the CMC was 25\% higher than 
the Milky Way value and this is may be due to the relatively large tracts of 
cold regions included in the average for the CMC.    
Of course,  variations in velocity dispersions between clouds could also 
contribute to the variation in their observed X-factors. However, we would 
expect similar velocity dispersions to characterize  gravitationally bound or 
virialized clouds of similar mass and size.  Moreover, since the average column 
densities of molecular clouds are also constant 
\citep[e.g.,][]{1981MNRAS.194..809L,2010A&A...519L...7L}, 
we would expect $<$\xcons$>$ $\sim$ $<$T$>^{-1}$ for such clouds.  
As recently pointed out by \citet{2013MNRAS.433.1223N},  
Galactic GMCs are characterized by very similar physical properties 
resulting in X-factors that span a relatively narrow range of values. 
They suggest that this is a result of the competition between stellar 
feedback and gravity that is necessary to maintain the relative stability of these objects.
Here we are suggesting that variation within that narrow range of 
observed \xco may be due to variations in the relative fractions of 
hot gas in these clouds. This hot gas, of course, is a direct product 
of the amount of stellar feedback the clouds are experiencing.  
Resolved maps of the internal X-factor distribution in additional clouds 
would be useful to determine the relative contribution of temperature to 
the derived values of \xco within GMCs in general and could directly test this idea.

\begin{figure}[htb!]
\epsscale{1.2}
\plotone{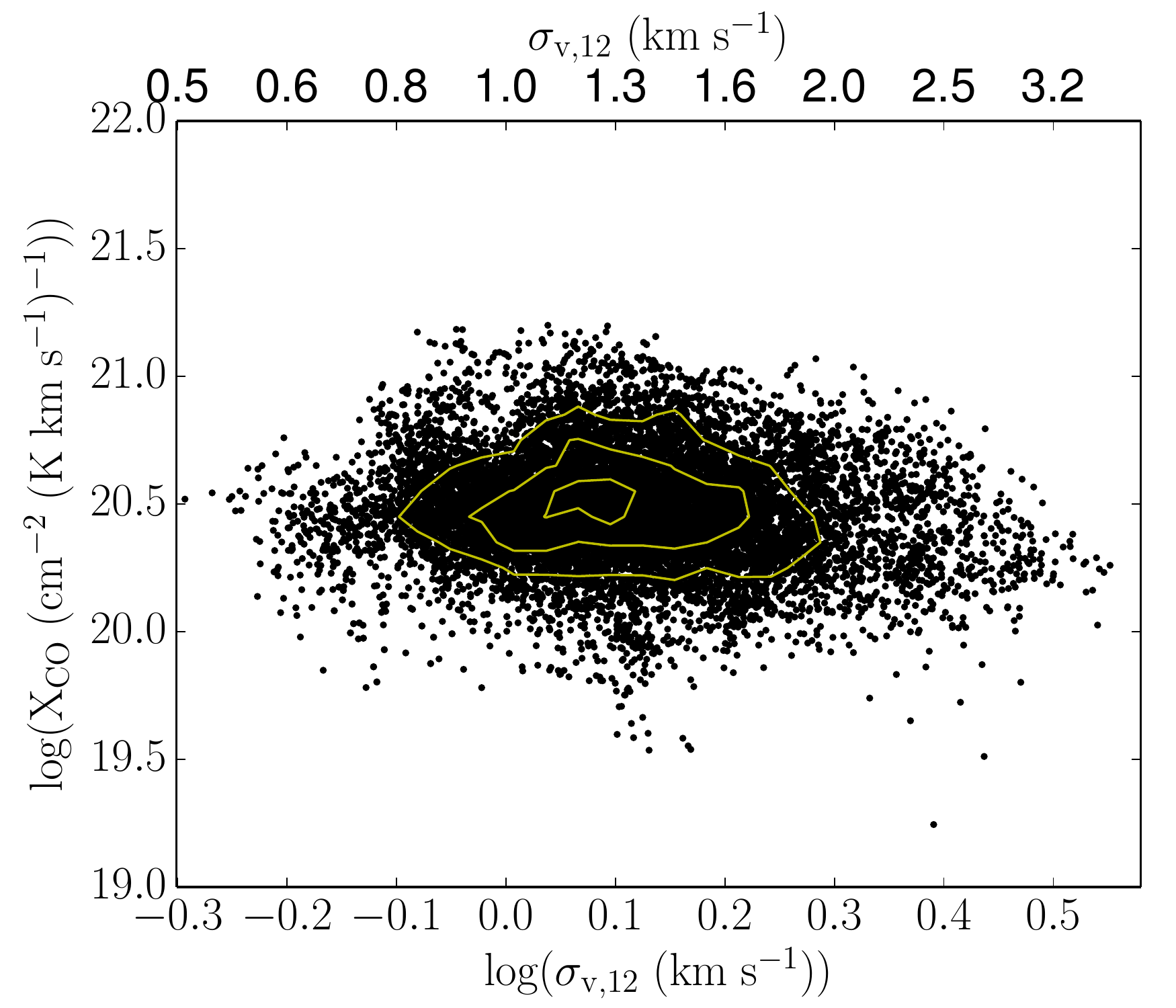}
\caption{\xcons-$\sigma_{v,12}$ relation for cold tiles, excluding tiles 11-14.
The yellow contours show the surface density of points.
There appears to be no strong correlation
between these two quantities. 
\label{fig:xcovssigma}}
\end{figure}

Absent of such additional data, it is instructive to compare our 
observations to recent simulations of turbulent cloud evolution 
that resolve synthetic clouds and derive the internal spatial 
distribution of the X-factor \citep{2011MNRAS.415.3253S,2011MNRAS.412.1686S}. 
These simulations are thus directly comparable to observations such as ours. 
The Shetty et al. simulations include chemical evolution and 
radiative transfer to enable predictions of CO line emission. 
However, the calculations do not include gas-dust chemical 
interactions and are thus not relevant to our earlier findings or 
discussion regarding [\thircons], [\ceions], and W(\thircons)/W(\ceions). 
But since \co is so optically thick we safely 
assume that effects of depletion/desorption are not important for 
our X-factor determinations and thus comparisons with these 
models are appropriate. Because the Shetty et al. simulations 
do not incorporate stellar feedback, the most appropriate comparison 
will be with our data in the regions of the CMC away from the embedded 
cluster (all tiles excluding 11-14). 
The simulated cloud of \citet{2011MNRAS.412.1686S} whose physical 
conditions match most closely those of typical GMCs is characterized 
by a global \xco  lower than we observe in the cold tiles of the 
CMC map (2.2 vs. 3.4  $\times$ 10$^{20}~{\rm cm}^{-2}~({\rm K~km~s}^{-1})^{-1}$) 
but closer to the Milky Way value. On sub-cloud scales 
\citet{2011MNRAS.412.1686S} find a correlation between \xco and \av 
(see their figure 4) similar to but somewhat weaker than we observe in the CMC. 
The simulations also predict that there should be a correlation between the 
X-factor and the velocity dispersion in the cloud, with \xco $\propto$ 
$\sigma_{V12}^{-0.5}$. In figure \ref{fig:xcovssigma} we plot \xco against 
the velocity dispersion, $\sigma_{V12}$ in the cold cloud regions\footnote{$\sigma_{V12}$ 
is the second moment of the \cons(2-1) profile $ \equiv \rm 
(\frac{\int T_{mb} (v-M1)^2 dv}{\int T_{mb}dv})^{\frac{1}{2}}$ where M1 is the 
first moment $\rm \int vT_{mb} dv$. $\sigma_{V12}$ equals the velocity dispersion if line profile 
is Gaussian.}. We find no correlation between these same quantities 
contrary to the predictions of the simulation. Although the simulations 
of individual clouds do not include any systematic temperature gradients 
or increases due to stellar heating, simulations were performed to  
investigate how increasing the overall cloud temperature can effect \xcons.  
These simulations predict a trend of decreasing X with T, consistent with 
what we observe here. The simulations find  \xco $\propto$ T$^{-0.5}$ 
for different temperature clouds which is only somewhat weaker than what is 
observed here, i.e., \xco $\propto$ T$^{-0.7}$ within the CMC.  
Overall, we conclude that the agreement between these simulations and 
the CMC is mixed and it is unclear whether any similarities between the 
simulations and the observations of the California cloud are more than coincidental.  

\section{Summary \& Conclusions}\label{sec:con}

In this study we presented the results of extinction and molecular-line 
mapping toward the most active star forming region within the nearby California GMC. 
The relationship between dust and gas properties in this active portion of the 
cloud was investigated with the following results:

1. 
The LTE molecular abundances 
of \thirco and \ceio in the cloud exhibit significant spatial variations. 
These variations are correlated with the spatial variation in gas (and presumably dust)  
temperature in the cloud. These temperature variations are likely caused by the 
heating from the adjacent, partially embedded,  LkH$\alpha$ 101 cluster.
We argue that the spatial variations in the derived \thirco and \ceio gas abundances are 
largely due to temperature dependent gas depletion/desorption on/off dust grains. 

2. The abundance ratio between \thirco and \ceio increases in lower extinction regions
 of the California cloud, suggesting selective photo-dissociation
of \ceio in those regions. This is likely caused by ambient UV radiation 
which apparently penetrates relatively deeply (i.e., \av $\la$ 15 mag) into the cloud, 
particularly near the embedded cluster. There is also evidence that the 
selective photo-dissociation of \ceio can suppress its abundance even 
in warm regions heated by the embedded cluster where evaporation off 
grain surfaces would otherwise lead to an increase in the \ceio abundance at these cloud depths. 

3. Dramatic spatial variations are also observed in the relationship between the 
\co integrated intensity, W$_{\rm ^{12}CO(2-1)}$, and \avns, 
particularly at the intermediate and high extinctions (\av $\ga$ 3-5 mag) 
primarily probed in this study. However, unlike the case for its two rarer isotopologues,
the variation in \co integrated intensity appears to be a direct result
of a spatial gradient  of the excitation temperature (and gas kinetic temperature) 
with extinction and not the result of any detectable variation in the \co abundance due to depletion/desorption effects. 

4. We compute the X-factor ($=$N(H$_2$)/W(CO)) for each individual pixel in 
our map and average the results to obtain $<$\xcons$>$ $=$ 2.53 $\times$ 
10$^{20}$~${\rm cm}^{-2}~({\rm K~km~s}^{-1})^{-1}$, a value somewhat higher than the Milky Way average.

5. On the sub-parsec scales in the CMC there is no single empirical value of the \co
 X-factor that can characterize the molecular gas in cold (\tk $\la$ 15 K) 
 regions with \av $\ga$ 3 magnitudes. For those regions we find that 
 \xco $\propto$ A$_{\rm V}^{0.7}$ with $<$\xcons$>$ $=$ 3.4 $\times$ 10$^{20}$ 
 ${\rm cm}^{-2}~({\rm K~km~s}^{-1})^{-1}$. We do however find a clear 
 correlation between W(\cons) and \av  in regions 
containing relatively hot (\tex $\ga$ 25 K ) molecular gas at \av $\ga$ 3 magnitudes
suggesting that, unlike the cold gas, the warm material may be characterized by a single value of \xcons.
However in this warm gas we find a value for the X-factor, \xco $=$ 1.5 
$\times$ 10$^{20}$~${\rm cm}^{-2}~({\rm K~km~s}^{-1})^{-1}$, significantly 
lower than  the averages for the cold gas,  the overall CMC and the Milky Way.

6. Overall we find an (inverse) correlation between \xco and \tex with \xco $\propto$ \texns$^{-0.7}$.  
Such a correlation may potentially explain the observed variations in global X-factors between 
GMCs as being due in large part to variations in the relative amounts of warm gas heated by OB stars within the clouds. 

\acknowledgments

We thank Kelsey Jorgenson for assistance with acquiring the CO observations 
and acknowledge useful discussions with Karin {\"O}berg. 
Suggestions from an anonymous referee led to numerous improvements in the paper. 
Carlos Roman acknowledges support from CONACYT project CB2010-152160, Mexico. 
This research is based in part on observations collected at the Centro 
Astron\'omico Hispano Alem\'an (CAHA) at Calar Alto, operated jointly by 
the Max-Planck-Institut f\"ur Astronomie and the Instituto de Astrof\'isica de Andaluc\'ia (CSIC). 
The Heinrich Hertz Submillimeter Telescope is operated by the Arizona Radio Observatory, 
which is part of Steward Observatory at The University of Arizona. 
This work was supported in part by National Science Foundation grant AST-1140030 to 
The University of Arizona.


\end{document}